\documentclass[preprint2]{aastex} 

\newcommand{\be}{\begin{equation}}
\newcommand{\ee}{\end{equation}}
\newcommand{\dd}{{\rm d}}
\newcommand{\gadget}{{\small GADGET}}
\newcommand{\gadgettwo}{{\small GADGET-2}}
\newcommand{\enzo}{{\small Enzo}}
\newcommand{\zeus}{{\small ZEUS}}
\newcommand{\ppm}{{\small PPM}}

\renewcommand{\vec}[1]{ {\bf #1} } 

\newcommand{\Lmax}{\ell_{\rm max}}
\newcommand{\rmax}{r_{\rm max}}
\newcommand{\Lbox}{L_{\rm box}}
\newcommand{\Nroot}{N_{\rm root}}

\newcommand{\Del}{\Delta}
\newcommand{\hinv}{{h^{-1}}}

\newcommand{\himpc}{\hinv{\rm\,Mpc}}
\newcommand{\hikpc}{\hinv{\rm\,kpc}}

\newcommand{\Msun}{\,\rm M_{\odot}}
\newcommand{\himsun}{\,\hinv\Msun}

\newcommand{\K}{\,\rm K}
\newcommand{\Le}{\Lbox/e}

\begin{document}


\title{Comparing AMR and SPH Cosmological Simulations:  
I.  Dark Matter \& Adiabatic Simulations}

\author{Brian W. O'Shea\altaffilmark{1,4}, Kentaro
Nagamine\altaffilmark{1, 2}, Volker Springel\altaffilmark{3},\\ 
Lars Hernquist\altaffilmark{2} \& Michael L. Norman\altaffilmark{1}}

\altaffiltext{1}{Center for Astrophysics and Space Sciences, 
University of California at San Diego, La Jolla, CA 92093, U.S.A.,  
Email:  bwoshea, mnorman@cosmos.ucsd.edu}

\altaffiltext{2}{Harvard-Smithsonian Center for Astrophysics,  
60 Garden St., Cambridge, MA 02138, U.S.A.,   
Email: knagamin, lars@cfa.harvard.edu}

 \altaffiltext{3}{Max-Planck-Institut f\"{u}r Astrophysik,  
Karl-Schwarzschild-Stra\ss{}e 1, 85740 Garching bei M\"{u}nchen,
Germany, Email: volker@mpa-garching.mpg.de}

\altaffiltext{4}{Theoretical Astrophysics (T-6), 
Los Alamos National Laboratory, Los Alamos, NM 87544, U.S.A.}


\begin{abstract}
We compare two cosmological hydrodynamic simulation codes in the
context of hierarchical galaxy formation: the Lagrangian 
smoothed particle
hydrodynamics (SPH) code `\gadget', and the Eulerian adaptive mesh
refinement (AMR) code `\enzo'. Both codes represent dark matter with
the N-body method but use different gravity solvers and
fundamentally different approaches for baryonic hydrodynamics.  The
SPH method in \gadget\ uses a recently developed `entropy
conserving' formulation of SPH, while for the mesh-based
\enzo\ two different formulations of Eulerian hydrodynamics are
employed: the piecewise parabolic method (\ppm) extended with a dual
energy formulation for cosmology, and the
artificial viscosity-based scheme used in the magnetohydrodynamics code
\zeus. In this paper we focus on a comparison of cosmological
simulations that follow either only dark matter, or also a
non-radiative (`adiabatic') hydrodynamic gaseous component.
We perform multiple simulations using both codes with varying spatial 
and mass resolution with identical initial conditions.

The dark matter-only runs agree generally quite well provided \enzo\
is run with a comparatively fine root grid and a low overdensity
threshold for mesh refinement, otherwise the abundance of low-mass
halos is suppressed. This can be readily understood as a consequence
of the hierarchical particle-mesh algorithm used by \enzo\ to compute
gravitational forces, which tends to deliver lower force resolution
than the tree-algorithm of \gadget\ at early times before any adaptive
mesh refinement takes place. At comparable force resolution we
find that the latter offers substantially better performance and lower
memory consumption than the present gravity solver in \enzo.

In simulations that include adiabatic gas dynamics we find general
agreement in the distribution functions of temperature, entropy, and
density for gas of moderate to high overdensity, as found inside dark
matter halos.  However, there are also some significant differences in
the same quantities for gas of lower overdensity.  For
example, at $z=3$ the fraction of cosmic gas that has temperature $\log T>
0.5$ is $\sim 80\%$ for both \enzo/\zeus\ and \gadget,
while it is $40-60\%$ for \enzo/\ppm. We argue that these
discrepancies are due to differences in the shock-capturing
abilities of the different methods.  In particular, we find that the
\zeus\ implementation of artificial viscosity in \enzo\
leads to some unphysical heating at early times in preshock
regions. While this is apparently a significantly weaker effect in \gadget, its
use of an artificial viscosity technique may also make it prone to some
excess generation of entropy which should be absent in
{\small ENZO/PPM}. Overall, the hydrodynamical results for \gadget\
are bracketed by those for \enzo/\zeus\ and \enzo/\ppm,
but are closer to \enzo/\zeus.

\end{abstract}

\keywords{galaxies: formation --- cosmology: theory --- methods: numerical}

\section{Introduction}\label{intro}

Within the currently leading theoretical model for structure
formation small fluctuations that were imprinted in the primordial
density field are amplified by gravity, eventually leading to
non-linear collapse and the formation of dark matter (DM) halos. Gas
then falls into the potential wells provided by the DM halos where it
is shock-heated and then cooled radiatively, allowing a fraction of
the gas to collapse to such high densities that star formation can
ensue. The formation of galaxies hence involves dissipative gas
dynamics coupled to the nonlinear regime of gravitational growth of
structure. The substantial difficulty of this problem is exacerbated
by the inherent three-dimensional character of structure formation in
a $\Lambda$CDM universe, where due to the shape of the primordial 
power spectrum a large range of wave modes becomes nonlinear in a very
short time, resulting in the rapid formation of objects with a wide 
range of masses which merge in geometrically complex ways into
ever more massive systems. Therefore, direct numerical simulations of 
structure formation which include hydrodynamics arguably provide the 
only method for studying this problem in its full generality.

Hydrodynamic methods used in cosmological simulations of galaxy
formation can be broken down into two primary classes: techniques
using an Eulerian grid, including `Adaptive Mesh Refinement' (AMR)
techniques, and those which follow the fluid elements in a Lagrangian
manner using gas particles, such as `Smoothed Particle Hydrodynamics'
(SPH).  Although significant amounts of work have been done on
structure/galaxy formation using both types of simulations, very few
detailed comparisons between the two simulation methods have been
carried out \citep[e.g.][]{Ka94,Frenk99}, despite the existence of 
widespread prejudices in the field with respect to alleged weaknesses 
and strengths of the different methods.

Perhaps the most extensive code comparison performed to date was the
{\em Santa Barbara cluster comparison project} \citep{Frenk99}, in
which several different groups ran a simulation of the formation of
one galaxy cluster, starting from identical initial conditions. They
compared a few key quantities of the formed cluster, such as
radially-averaged profiles of baryon and dark matter density, gas
temperature and X-ray luminosity.  Both Eulerian (fixed grid and AMR)
and SPH methods were used in this study. Although most of the measured
properties of the simulated cluster agreed reasonably well between
different types of simulations (typically within $\sim 20\%$), there
were also some noticeable differences which largely remained unexplained,
for example in the central entropy profile, or in the baryon fraction
within the virial radius.  Later simulations by \citet{Ascasibar} compare
results from the Eulerian AMR code ART \citep{Kravtsov02} with the 
entropy-conserving version of \gadget.  They find that the entropy-conserving
version of \gadget\ significantly improves agreement with grid codes when
examining the central entropy profile of a galaxy cluster, though the results
are not fully converged.  The \gadget\
result using the new hydro formulation now shows an entropy floor -- 
in the Santa Barbara paper the SPH codes typically did not display any 
trend towards a floor in entropy at the center of the cluster while 
the grid-based codes generally did.
The ART code produces results that agree 
extremely well with the grid codes used in the comparison. The observed
convergence in cluster properties is encouraging, but there is still 
a need to explore other systematic differences between simulation methods.

The purpose of the present study is to compare two different types of
modern cosmological hydrodynamic methods, SPH and AMR, in greater
depth, with the goal of obtaining a better understanding of the
systematic differences between the different numerical
techniques. This will also help to arrive at a more reliable
assessment of the systematic uncertainties in present numerical
simulations, and provide guidance for future improvements in numerical
methods.  The codes we use are
`\gadget'\footnote{http://www.MPA-Garching.MPG.DE/gadget/}, an SPH
code developed by \citet*{Springel00}, 
and `\enzo'\footnote{http://www.cosmos.ucsd.edu/enzo/}, 
an AMR code
developed by \citet{Bryan97, Bryan99}.  In this paper, we focus our
attention on the clustering properties of dark matter and on the
global distribution of the thermodynamic quantities of cosmic gas,
such as temperature, density, and entropy of the gas.  Our work is
thus complementary to the Santa Barbara cluster comparison project
because we examine cosmological volumes that include many halos and
a low-density intergalactic medium, rather than focusing on a single
particularly well-resolved halo. We also include 
idealized tests designed to highlight the effects of artificial
viscosity and cosmic expansion.

The present study is the first paper in a series that aims at
providing a comprehensive comparison of AMR and SPH methods applied to
the dissipative galaxy formation problem.  In this paper,
we describe the general code methodology, and present
fundamental comparisons between dark matter-only runs and runs that
include ordinary `adiabatic' hydrodynamics.  This paper is meant to 
provide statistical comparisons between simulation codes, and we leave
detailed comparisons of baryon properties in individual halos to a later
paper. Additionally, we plan to compare
simulations using radiative cooling, star formation, and supernova
feedback in forthcoming work.

The organization of this paper is as follows. We provide a short
overview of the two codes being compared in Section~\ref{code}, and
then describe the details of our simulations in Section~\ref{sims}.
Our comparison is then conducted in two steps.  We first compare the
dark matter-only runs in Section~\ref{dm} to test the gravity solver
of each code. This is followed in Section~\ref{baryon} with a detailed
comparison of hydrodynamic results obtained in adiabatic cosmological
simulations.  We then discuss effects of artificial viscosity in
Section~\ref{viscosity}, and the timing and memory usage of the two
codes in Section~\ref{timing}. Finally, we conclude in
Section~\ref{discussion} with a discussion of our findings.


\section{Code details}\label{code}

\subsection{Adaptive mesh refinement code}\label{amr}

`\enzo' is an adaptive mesh refinement cosmological simulation code
developed by Bryan et al. \citep{Bryan97, Bryan99, Norman99, Bryan01}. 
This code couples an N-body particle-mesh (PM) solver 
\citep[e.g.][]{Efstathiou85, Hockney88} used to follow the 
evolution of collisionless dark matter with an Eulerian AMR method for 
ideal gas dynamics by \citet{Berger89}, which allows extremely high
 dynamic range in gravitational physics and hydrodynamics 
in an expanding universe.  

Unlike moving mesh methods \citep{Pen, Gnedin} or methods that subdivide 
individual cells \citep{Adjerid}, Berger \& Collela's AMR (also referred 
to as \emph{structured} AMR) utilizes an adaptive hierarchy of grid 
patches at varying levels of resolution.  Each rectangular grid patch 
(referred to as a ``grid'') covers some region of space in its 
\emph{parent grid} which requires higher resolution, and can itself 
become the parent grid to an even more highly resolved \emph{child grid}. 
ENZO's implementation of structured AMR places no restrictions on 
the number of grids at a given level of refinement, or on the number of 
levels of refinement.  However, owing to limited computational resources 
it is practical to institute a maximum level of refinement $\Lmax$.  

The \enzo\ implementation of AMR allows arbitrary integer ratios of parent and
child grid resolution.  However, in this study we choose to have a
refinement factor of two, meaning that each child grid has a factor of
two higher spatial resolution than its parent grid.  The ratio of
boxsize to the maximum grid resolution of a given simulation is
therefore $L/e = \Nroot \times 2^{\Lmax}$, where $\Nroot$ is the
number of cells along one edge of the root grid, and $\Lmax$ is the
maximum level of refinement allowed.

The AMR grid patches are the primary data structure in \enzo.  Each
patch is treated as an individual object which can contain both field
variables and particle data.  Individual grids are organized into a 
dynamic, distributed hierarchy of mesh patches.  Every processor
keeps a description of the entire grid hierarchy at all times, so that 
each processor knows where all grids are.  However, baryon and particle 
data for a given grid only exists on a single processor.  The code 
handles load balancing 
on a level-by-level basis such that the workload on each level is 
distributed uniformly across all processors.  The MPI message passing 
library is used to transfer data between processors.

Each grid patch in \enzo\ contains arrays of values for baryon and 
particle quantities.  The baryon quantities are stored in arrays 
with the dimensionality of the simulation itself, which can
 be 1, 2 or 3 spatial dimensions.  Grids are 
partitioned into a core of \emph{real zones} and a surrounding layer 
of \emph{ghost zones}.  The real zones store field values and ghost 
zones are used to temporarily store values which have been obtained 
directly from neighboring grids or interpolated from a parent grid.  
These zones are necessary to accommodate the computational stencil of 
the hydrodynamics solvers (Sections~\ref{ppmhydro} and~\ref{zeushydro}) 
and the gravity solver (Section~\ref{enzo-grav}).  
The hydro solvers require ghost zones which 
are three cells deep and the gravity solver requires 6 ghost zones 
on every side of the real zones.  This can lead to significant memory 
and computational overhead, particularly for smaller grid patches
at high levels of refinement.

Since the addition of more highly refined grids is adaptive,  
the conditions for refinement must be specified.  The criteria of 
refinement can be set by the threshold value of the overdensity of 
baryon gas or dark matter in a cell (which is really a refinement on
the mass of gas or DM in a cell), the local Jeans length, the 
local density gradient, or local pressure and energy gradients.  
A cell reaching any or all of these criteria 
will then be flagged for refinement.  
Once all cells of given level have been flagged, rectangular boundaries 
are determined which minimally encompass them. A refined grid patch is 
introduced within each such bounding rectangle. Thus, the cells needing 
refinement as well as adjacent cells within the patch which do not need 
refinement are refined. While this approach is not as memory efficient as 
cell-splitting AMR schemes, it offers more freedom with 
finite difference stencils. For example, PPM requires a stencil of seven 
cells per dimension. This cannot easily be accommodated in cell-splitting 
AMR schemes. 
In the current study we use baryon and 
dark matter overdensities as our refinement criteria.

Two different hydrodynamic methods are implemented in \enzo: the
piecewise parabolic method (PPM) developed by
\citet{Woodward84} and extended to cosmology by \citet{Bryan95}, 
and the hydrodynamic method used in the
magnetohydrodynamics (MHD) code \zeus. Below, we describe both of
these methods in turn, noting that PPM is viewed as the preferred
choice for cosmological simulations since it is higher-order-accurate
and is based on a technique that does not require artificial
viscosity to resolve shocks.  Additional physical processes (such as radiative
cooling and star formation and feedback) are also implemented within \enzo\ but
 are outside the scope of the present comparison study and will not
be discussed here.

In \enzo, resolution of the equations being solved is adaptive in time 
as well as in space.  The timestep in \enzo\ is satisfied on a level-by-level
 basis by finding the largest timestep such that the Courant condition (and 
an analogous condition for the dark matter particles) is satisfied by every 
cell on that level.  All cells on a given level are advanced using the same 
timestep.  Once a level $L$ has been advanced in time by $\Delta t_L$, all 
grids at level $L+1$ are advanced, using the same criteria for timestep 
calculation described above, until they reach the same physical time as the 
grids at level $L$.  At this point grids at level $L+1$ exchange flux 
information with their parents grids, providing a more accurate solution 
on level L.  This step, controlled by the parameter \emph{FluxCorrection} in
\enzo, is extremely important, and can significantly affect simulation results
if not used in an AMR calculation.
At the end of every timestep on every level each grid updates 
its ghost zones by exchanging information with its neighboring grid patches 
(if any exist) and/or by interpolating from a parent grid.  
In addition, cells are examined to see if they should be refined or 
de-refined, and the entire grid hierarchy is rebuilt at that 
level (including all more highly refined levels).  The timestepping and 
hierarchy advancement/rebuilding process described here is repeated 
recursively on every level to the specified maximum level of 
refinement in the simulation.


\subsubsection{Hydrodynamics with the piecewise parabolic method}
\label{ppmhydro}

The primary hydrodynamic method used in \enzo\ is based on the
piecewise parabolic method (PPM) of \citet{Woodward84} which has been
significantly modified for the study of cosmological fluid flows.  The
method is described in \citet{Bryan95}, but we provide a short
description here for clarity.  PPM is a higher order-accurate version of
Godunov's method for ideal gas dynamics with third order-accurate piecewise parabolic
monotonic interpolation and a nonlinear Riemann solver for shock
capturing.  It does an excellent job of capturing strong shocks in at
most two cells.  Multidimensional schemes are built up by directional
splitting and produce a method that is formally second order-accurate
in space and time which explicitly conserves mass, linear momentum, and energy.  
The conservation laws for fluid mass, momentum and energy
density are written in comoving coordinates for a
Friedman-Robertson-Walker space-time. Both the conservation laws and
the Riemann solver are modified to include gravity, which is solved
using the particle-mesh (PM) technique (see Section~\ref{enzo-grav}).  
Unlike the other hydrodynamic
techniques used in this comparison, PPM does not use artificial
viscosity.  This is an important methodological difference compared
with other methods that use artificial viscosity, such as the \zeus\ 
hydrodynamics algorithm and \gadget 's SPH method.

In order to more accurately treat situations involving bulk hypersonic
motion, where the kinetic energy of the gas can dominate the internal
energy by many orders of magnitude, both the gas internal energy
equation and total energy equation are solved everywhere on the grid
at all times.  This \emph{dual energy formulation} ensures that the
method produces the correct entropy jump at strong shocks and also
yields accurate pressures and temperatures in cosmological hypersonic
flows.


\subsubsection{The \zeus\ hydrodynamic algorithm}
\label{zeushydro}

As a check on PPM, \enzo\ also includes an implementation of the
finite-difference hydrodynamic algorithm employed in the compressible 
magnetohydrodynamics code `\zeus'
\citep{Stone92a, Stone92b}.  Fluid transport is solved on a Cartesian
grid using the upwind, monotonic advection scheme of \citet{Leer77} 
within a multistep 
(operator split) solution procedure which is fully explicit in time.  
This method is formally second order-accurate in space but 
first order-accurate in time.

Operator split methods break
the solution of the hydrodynamic equations into parts, with each part
representing a single term in the equations.  Each part is evaluated
successively using the results preceding it.  The individual parts of
the solution are grouped into two steps, called the \emph{source} and
\emph{transport} steps.

The \zeus\ method uses a von Neumann-Richtmyer artificial viscosity 
to smooth shock discontinuities that may
appear in fluid flows and can cause a break-down of finite-difference
equations.  The artificial viscosity term is added in the source terms
as
\begin{eqnarray}
\rho \frac{\partial\textbf{v}}{\partial t} &=& - \nabla p - \rho \nabla \phi - \nabla \cdot \textbf{Q} \\
\frac{\partial e}{\partial t} &=& -p \nabla \cdot \textbf{v} - \textbf{Q} : \nabla \textbf{v}, 
\end{eqnarray}
where \textbf{v} is the baryon velocity, $\rho$ is the mass density,
$p$ is pressure, $e$ is internal energy density of gas and \textbf{Q}
is the artificial viscosity stress tensor, such that:

\begin{eqnarray}
Q_{ii} = \left\{ \begin{array}{ll}
Q_{\rm AV} \rho_b  (a \Delta v_{i} + \dot{a} \Delta x_i)^2, 
& \textrm{for $a \Delta v_{i} + \dot{a} \Delta x_i < 0$}  \\
0 & \textrm{otherwise}\\
\end{array} \right. 
\end{eqnarray}
and
\begin{equation}
Q_{ij} = 0  \;\;\;{\rm for}\;\; i \ne j .
\end{equation}

$\Delta x_i$ and $\Delta v_{i}$ refer to the comoving 
width of the grid cell
along the $i$-th axis and the corresponding difference in gas
peculiar velocities across the grid cell, respectively, and $a$ is the
cosmological scale factor.  $Q_{\rm AV}$ is a constant with a typical
value of 2. We refer the interested reader to \citet{Anninos} for more details.

The limitation of a technique that uses an artificial viscosity is that,
while the correct Rankine-Hugoniot jump conditions are achieved,
shocks are broadened over 6-8 mesh cells and are thus not treated as true
discontinuities. This may cause unphysical pre-heating of gas upstream
of the shock wave, as discussed in \citet{Anninos}.

\subsubsection{The \enzo\ Gravity Solver}\label{enzo-grav}

There are multiple methods to compute the gravitational potential 
(which is an elliptic equation in the Newtonian limit) in a structured 
AMR framework.  One way would be to model the dark matter (or other
collisionless particle-like objects, such as stars) as a second fluid 
in addition to the baryonic fluid and solve the collisionless
Boltzmann equation, which follows the evolution of the fluid density in
six-dimensional phase space.  However, this is computationally prohibitive 
owing to the large dimensionality of the problem, making this approach 
unappealing for the cosmological AMR code.

Instead, \enzo\ uses a particle-mesh N-body method to calculate 
the dynamics of collisionless systems.  This method follows trajectories 
of a representative sample of individual particles and is much more 
efficient than a direct solution of the Boltzmann equation in most 
astrophysical situations. 
The gravitational potential is computed by solving the elliptic 
Poisson's equation:
\begin{equation}
\nabla^2 \phi = 4\pi G \rho,
\end{equation}
where $\phi$ is the gravitational potential and 
$\rho$ is the density of both the collisional fluid (baryonic gas)
and the collisionless fluid (dark matter particles).

These equations are finite-differenced and for simplicity are solved with the
same timestep as the hydrodynamic equations.  This universal timestep is taken
to be the minimum of the baryon and dark matter timesteps, thus ensuring
numerical stability.  The dark matter particles are distributed onto the grids
using the cloud-in-cell (CIC) interpolation technique to form a spatially
discretized density field (analogous to the baryon densities used to calculate
the equations of hydrodynamics).  After sampling dark matter density onto the
grid and adding baryon density if it exists (to get the total matter density
in each cell), the gravitational potential is calculated on the periodic root
grid using a fast Fourier transform.  In order to calculate more accurate
potentials on subgrids, \enzo\ re-samples the dark matter density onto
individual subgrids using the same CIC method as on the root grid, but at
higher spatial resolution (and again adding baryon densities if applicable).
Boundary conditions are then interpolated from the potential values on the
parent grid (with adjacent grid patches on a given level communicating to
ensure that their boundary values are the same), and then a multigrid
relaxation technique is used to calculate the gravitational potential at every
point within a subgrid.  Forces are computed on the mesh by
finite-differencing potential values and are then interpolated to the particle
positions, where they are used to update the particle's position and velocity
information.  Potentials on child grids are computed recursively and particle
positions are updated using the same timestep as in the hydrodynamic
equations.  Particles are stored in the most highly refined grid patch at the
point in space where they exist, and particles which move out of a subgrid
patch are sent to the grid patch covering the adjacent volume with the finest
spatial resolution, which may be of the same spatial resolution, coarser, or finer
than the grid patch that the particles are moved from.  This takes place in a
communication process at the end of each timestep on a level.

At this point it is useful to emphasize that the effective force
resolution of an adaptive particle-mesh calculation is approximately twice
as coarse as the grid spacing at a given level of resolution.  The 
potential is solved in each grid cell;
however, the quantity of interest, namely the acceleration, is the gradient
of the potential, and hence two potential values are required to calculate
this.  In addition, it should be noted that the adaptive particle-mesh
technique described here is very memory intensive: in order to ensure accurate 
force resolution at grid edges the multigrid relaxation
method used in the code requires a layer of ghost zones which is very deep --
typically 6 cells in every direction around the edge of a grid patch.  This greatly
adds to the memory requirements of the simulation, particularly because subgrids
are typically small (on the order of $12^3 - 16^3$ real cells for a standard 
cosmological calculation) and ghost zones can dominate the memory and computational 
requirements of the code.


\subsection{Smoothed particle hydrodynamics code}\label{sph}

In this study, we compare \enzo\ with a new version of the parallel TreeSPH
code \gadget\ (Springel 2005, in preparation), which combines smoothed
particle hydrodynamics with a hierarchical tree algorithm for gravitational
forces.  Codes with a similar principal design
\citep[e.g.][]{He89,Na93,Ka96,Da97} have been employed in cosmology for a
number of years.  Compared with previous SPH implementations, the new version
\gadgettwo\ used here differs significantly in its formulation of SPH (as
discussed below), in its timestepping algorithm, and in its parallelization
strategy. In addition, the new code optionally allows the computation of
long-range forces with a particle-mesh (PM) algorithm, with the tree algorithm
supplying short-range gravitational interactions only.  This `TreePM' method
can substantially speed up the computation while maintaining the large dynamic
range and flexibility of the tree algorithm.

\subsubsection{Hydrodynamical method}

SPH uses a set of discrete tracer particles to describe the state of a fluid,
with continuous fluid quantities being defined by a kernel interpolation
technique if needed \citep{Lu77,Gi77,Mo92}. The particles with coordinates
$\vec{r}_i$, velocities $\vec{v}_i$, and masses $m_i$ are best thought of as
fluid elements that sample the gas in a Lagrangian sense. The thermodynamic
state of each fluid element may either be defined in terms of its thermal
energy per unit mass, $u_i$, or in terms of the entropy per unit mass, $s_i$.
We in general prefer to use the latter as the independent thermodynamic
variable evolved in SPH, for reasons discussed in full detail by \citet{SH02}.
In essence, use of the entropy allows SPH to be formulated so that both
energy and entropy are manifestly conserved, even when adaptive smoothing
lengths are used \citep[see also][]{He93}.   In the following we
summarize the ``entropy formulation'' of SPH, which is implemented
in \gadgettwo~as suggested by \citet{SH02}.

We begin by noting that it is more convenient to work with an entropic
function defined by $A \equiv P/\rho^\gamma$, instead of directly using the
thermodynamic entropy $s$ per unit mass.  Because $A=A(s)$ is only a function
of $s$ for an ideal gas, we will simply call $A$ the `entropy' in what
follows. Of fundamental importance for any SPH formulation is the density
estimate, which \gadget\ calculates in the form \be \rho_i = \sum_{j=1}^N m_j
W(|\vec{r}_{ij}|,h_i),
\label{eqndens}\ee where $\vec{r}_{ij}\equiv \vec{r}_i - \vec{r}_j$, and
$W(r,h)$ is the SPH smoothing kernel. In the entropy formulation of the code,
the adaptive smoothing lengths $h_i$ of each particle are defined such that
their kernel volumes contain a constant mass for the estimated density;
i.e.~the smoothing lengths and the estimated densities obey the (implicit)
equations \be \frac{4\pi}{3} h_i^3 \rho_i = N_{\rm sph}\overline{m},
\label{eqhsml}\ee where $N_{\rm sph}$ is the typical number of smoothing
neighbors, and $\overline{m}$ is the average particle mass. Note that in
traditional formulations of SPH, smoothing lengths are typically chosen such
that the number of particles inside the smoothing radius $h_i$ is equal to a
constant value $N_{\rm sph}$.

Starting from a discretized version of the fluid Lagrangian, one can show
\citep{SH02} that the equations of motion for the SPH particles are given by
\be \frac{\dd \vec{v}_i}{\dd t} = - \sum_{j=1}^N m_j \left[ f_i
  \frac{P_i}{\rho_i^2} \nabla_i W_{ij}(h_i) + f_j \frac{P_j}{\rho_j^2}
  \nabla_i W_{ij}(h_j) \right],
\label{eqnmot} 
\ee where the coefficients $f_i$ are defined by \be f_i = \left[ 1 +
  \frac{h_i}{3\rho_i}\frac{\partial \rho_i}{\partial h_i} \right]^{-1} \, ,
\ee and the abbreviation $W_{ij}(h)= W(|\vec{r}_{i}-\vec{r}_{j}|, h)$ has been
used. The particle pressures are given by $P_i=A_i \rho_i^{\gamma}$. Provided
there are no shocks and no external sources of heat, the equations above
already fully define reversible fluid dynamics in SPH. The entropy $A_i$ of
each particle simply remains constant in such a flow.

However, flows of ideal gases can easily develop discontinuities
where entropy must be generated by microphysics. Such shocks need to
be captured by an artificial viscosity technique in SPH. To this end
\gadget\ uses a viscous force \be \left. \frac{\dd \vec{v}_i}{\dd
t}\right|_{\rm visc.}  = -\sum_{j=1}^N m_j \Pi_{ij}
\nabla_i\overline{W}_{ij} \, .
\label{eqnvisc}
\ee For the simulations of this paper, we use a standard Monaghan-Balsara
artificial viscosity $\Pi_{ij}$ \citep{Mo83,Bal95}, parameterized in the
following form: \be
\label{eqvisc}
\Pi_{ij}=\left\{
\begin{tabular}{cl}
${\left[-\alpha c_{ij} \mu_{ij} +2\alpha \mu_{ij}^2\right]}/{\rho_{ij}}$ & \mbox{if
$\vec{v}_{ij}\cdot\vec{r}_{ij}<0$} \\
0 & \mbox{otherwise}, 
\label{sphviscosity}\\
\end{tabular}
\right.  \ee with \be \mu_{ij}=\frac{h_{ij}\,\vec{v}_{ij}\cdot\vec{r}_{ij} }
{\left|\vec{r}_{ij}\right|^2}. \ee Here $h_{ij}$ and $\rho_{ij}$ denote
arithmetic means of the corresponding quantities for the two particles $i$ and
$j$, with $c_{ij}$ giving the mean sound speed.  The symbol
$\overline{W}_{ij}$ in the viscous force is the arithmetic average of the two
kernels $W_{ij}(h_i)$ and $W_{ij}(h_j)$.  The strength of the viscosity is
regulated by the parameter $\alpha$, with typical values in the range
$0.75-1.0$.  Following \citet{St96}, \gadget\ also uses an additional
viscosity-limiter in Eqn.~(\ref{eqvisc}) in the presence of strong shear flows
to alleviate angular momentum transport.

Note that the artificial viscosity is only active when fluid elements
approach one another in physical space, to prevent particle
interpenetration.  In this case, entropy is generated by the viscous
force at a rate \be \frac{\dd A_i}{\dd t} =
\frac{1}{2}\frac{\gamma-1}{\rho_i^{\gamma-1}}\sum_{j=1}^N m_j \Pi_{ij}
\vec{v}_{ij}\cdot\nabla_i \overline{W}_{ij} \,,
\label{entropyproduction}\ee transforming kinetic energy of gas motion
irreversibly into heat.

We have also run a few simulations with a `conventional formulation' of SPH
in order to compare its results with the `entropy formulation'. This
conventional formulation is characterized by the following
differences.  Equation~(\ref{eqhsml}) is replaced by a choice of
smoothing length that keeps the number of neighbors constant. In the
equation of motion, the coefficients $f_i$ and $f_j$ are always equal
to unity, and finally, the entropy is replaced by the thermal energy
per unit mass as an independent thermodynamic variable. The thermal
energy is then evolved as \be \frac{\dd u_i}{\dd t} = \sum_{j=1}^N m_j
\left( \frac{P_i}{\rho_i^2} + \frac{1}{2}\Pi_{ij}\right) \vec{v}_{ij}
\cdot \nabla_i \overline{W}_{ij}, \ee with the particle pressures
being defined as $P_i=(\gamma-1)\rho_i u_i$.

\subsubsection{Gravitational method}
\label{gravmethod}

In the \gadget\ code, both the collisionless dark matter and the
gaseous fluid are represented by particles, allowing the self-gravity
of both components to be computed with gravitational N-body
methods. Assuming a periodic box of size $L$, the forces can be
formally computed as the gradient of the periodic peculiar potential
$\phi$, which is the solution of \be \nabla^2 \phi(\vec{x}) = 4\pi G
\sum_i m_i  \left[  - \frac{1}{L^3} +  \sum_{\vec{n}} 
\tilde\delta(\vec{x}-\vec{x}_i-\vec{n}L)\right],
\label{eqnpecpot}\ee where the sum over $\vec{n}=(n_1, n_2, n_3)$ extends 
over all integer triples.  The function $\tilde\delta(\vec{x})$ is a
normalized softening kernel, which distributes the mass of a point-mass
over a scale corresponding to the gravitational softening length
$\epsilon$. The \gadget\ code adopts the spline kernel used in SPH for 
$\tilde\delta(\vec{x})$, with a scale length chosen such that the 
force of a point mass becomes fully Newtonian at a separation of 
$2.8\,\epsilon$, with a gravitational potential at zero lag equal to 
$-Gm/\epsilon$, allowing the interpretation of $\epsilon$ as a Plummer 
equivalent softening length.

Evaluating the forces by direct summation over all particles becomes rapidly
prohibitive for large $N$ owing to the inherent $N^2$ scaling of this approach.
Tree algorithms such as that used in \gadget\ overcome this problem by using a
hierarchical multipole expansion in which distant particles are grouped into
ever larger cells, allowing their gravity to be accounted for by means of a
single multipole force. Instead of requiring $N-1$ partial forces per
particle, the gravitational force on a single particle can then be computed
from just ${\cal O}({\rm log} N)$ interactions.

It should be noted that the final result of the tree algorithm will in
general only represent an approximation to the true force described by
Eqn.~(\ref{eqnpecpot}). However, the error can be controlled
conveniently by adjusting the opening criterion for tree nodes, and,
provided sufficient computational resources are invested, the tree
force can be made arbitrarily close to the well-specified correct
force.

The summation over the infinite grid of particle images required for
simulations with periodic boundary conditions can also be treated in the tree
algorithm. \gadget\ uses the technique proposed by \citet{He91} for this
purpose.  Alternatively, the new version \gadgettwo used in this study allows
the pure tree algorithm to be replaced by a hybrid method consisting of a
synthesis of the particle-mesh method and the tree algorithm. \gadget's
mathematical implementation of this so-called TreePM method
\citep{Xu95,Bode2000,Bagla02a} is similar to that of \citet{Bagla02b}.  The
potential of Eqn.~(\ref{eqnpecpot}) is explicitly split in Fourier space into
a long-range and a short-range part according to $\phi_{\vec{k}} = \phi^{{\rm
    long}}_{\vec{k}} + \phi^{{\rm short}}_{\vec{k}}$, where \be \phi^{{\rm
    long}}_{\vec{k}} = \phi_{\vec{k}} \exp(-\vec{k}^2 r_s^2), \ee with $r_s$
describing the spatial scale of the force-split.  This long range potential
can be computed very efficiently with mesh-based Fourier methods. Note that if
$r_s$ is chosen slightly larger than the mesh scale, force anisotropies that
exist in plain PM methods can be suppressed to essentially arbitrarily small
levels.

The short range part of the potential can be solved in real space by
noting that for $r_s \ll L$ the short-range part of the potential is
given by \be \phi^{{\rm short}}(\vec{x}) = - G \sum_i \frac{m_i}{r_i}
{\rm erfc}\left(\frac{r_i} {2 r_s}\right).
\label{eqshortfrc}\ee Here $r_i = \min(|\vec{x}-\vec{r}_i - \vec{n}L|)$ is
defined as the smallest distance of any of the images of particle $i$
to the point $\vec{x}$.  The short-range force can  still be
computed by the tree algorithm, except that the force law is modified
according to Eqn.~(\ref{eqshortfrc}). However, the tree only needs to
be walked in a small spatial region around each target particle
(because the complementary error function rapidly falls for $r >
r_s$), and no corrections for periodic boundary conditions are
required, which together can result in a very substantial performance
gain. One typically also gains accuracy in the long range force, which
is now basically exact, and not an approximation as in the tree
method. In addition, the TreePM approach maintains all of the most
important advantages of the tree algorithm, namely its insensitivity
to clustering, its essentially unlimited dynamic range, and its
precise control about the softening scale of the gravitational force.


\section{The simulation set}
\label{sims}

\notetoeditor{Put Table 1 here.}

\notetoeditor{Put Table 2 here.}

In all of our simulations, we adopt the standard concordance cold dark
matter model of a flat universe with $\Omega_m=0.3$,
$\Omega_{\Lambda}=0.7$, $\sigma_8=0.9$, $n=1$, and $h=0.7$.  For
simulations including hydrodynamics, we take the baryon mass density
to be $\Omega_b=0.04$. The simulations are initialized at redshift
$z=99$ using the \citet{E-Hu} transfer function. For the dark
matter-only runs, we chose a periodic box of comoving size
$12\,\himpc$, while for the adiabatic runs we preferred $3\, \himpc$ to
achieve higher mass resolution, although the exact size of the
simulation box is of little importance for the present
comparison. Note however that this is different in simulations that
also include cooling, which imprints additional physical scales.  We
place the unperturbed dark matter particles at the vertices of a
Cartesian grid, with the gas particles offset by half the mean
interparticle separation in the \gadget\ simulations. These particles
are then perturbed by the Zel'dovich approximation for the initial 
conditions. In \enzo, fluid elements are represented by the values 
at the center of the cells and are also perturbed using the Zel`dovich
approximation.

For both codes, we have run a large number of simulations, varying the
resolution, the physics (dark matter only, or dark matter with
adiabatic hydrodynamics), and some key numerical parameters.  Most of
these simulations have been evolved to redshift $z=3$.  We give a full
list of all simulations we use in this study in
Tables~\ref{table.sphsim} and \ref{table.amrsim} for \gadget\ and
\enzo, respectively; below we give some further explanations for
this simulation set.

We performed a suite of dark matter-only simulations in order to
compare the gravity solvers in \enzo\ and \gadget.  For \gadget, the
spatial resolution is determined by the gravitational softening length
$\epsilon$, while for \enzo\ the equivalent quantity is given by the
smallest allowed mesh size $e$ (note that in \enzo\ the gravitational
force resolution is approximately twice as coarse as this:  
see Section~\ref{enzo-grav}). 
Together with the boxsize $\Lbox$, we can then define a dynamic range 
$\Le$ to characterize a simulation (for simplicity we use $\Le$ for 
\gadget\ as well instead of $\Lbox/\epsilon$).
For our basic set of runs with $64^3$ dark matter particles we varied
$\Le$ from 256 to 512, 1024, 2048 and 4096 in \enzo. We also computed
corresponding \gadget\ simulations, except for the $\Le=4096$ case,
which presumably would already show sizable two-body scattering
effects. Note that it is common practice to run collisionless tree
N-body simulations with softenings in the range $1/25 - 1/30$ of the
mean interparticle separation, translating to $\Le =1600-1920$ for a
$64^3$ simulation.

Unlike in \gadget, the force accuracy in \enzo\ at early times also depends on the root grid
size. For most of our runs we used a root grid with $64^3$ cells, 
but we also performed \enzo\ runs with a $128^3$ root grid
in order to test the effect of the root grid size on the dark matter halo mass
function.  Both $64^3$ and $128^3$ particles were used, with the 
number of particles never exceeding the size of the root grid.

Our main interest in this study lies, however, in our second set of
runs, where we additionally follow the hydrodynamics of a baryonic
component, modeled here as an ideal, non-radiative gas.  As above, we
use $64^3$ DM particles and $64^3$ gas particles (for \gadget), or a
$64^3$ root grid (for \enzo), in most of our runs, though as before we
also perform runs with $128^3$ particles and root grids. Again, we vary the
dynamic range $\Le$ from 256 to 4096 in \enzo, and parallel this with
corresponding \gadget\ runs, except for the $\Le=4096$ case.

An important parameter of the AMR method is the mesh-refinement criterion.
Usually, \enzo\ runs are configured such that grid refinement occurs when the
dark matter mass in a cell reaches a factor of 4.0 times the mean dark matter
mass expected in a cell at root grid level, or if it has a factor of 8.0 times
the mean baryonic mass of a root level cell, but several runs were performed
with a threshold density set to 0.5 of the standard values for both dark
matter and baryon density.  All that the ``refinement overdensity'' criteria
does is set the maximum gas or dark matter mass which may exist in a given cell 
before that cell must be refined based on a multiple of the mean cell mass on 
the root grid.  For example, a baryon overdensity threshold of 4.0 means 
that a cell is forced to refine once a 
cell has accumulated more than 4 times the mean cell mass on the root grid.

When the refinement overdensity is set to the
higher value discussed here, the simulation may fail to properly identify
small density peaks at early times, so that they are not well resolved by
placing refinements on them. As a result, the formation of low-mass DM halos
or substructure in larger halos may be suppressed.  Note that lowering the
refinement threshold results in a significant increase in the number of
refined grids, and hence a significant increase in the computational cost of a
simulation; i.e., one must tune the refinement criteria to compromise between
performance and accuracy.

We also performed simulations with higher mass and spatial resolution, ranging
up to $2\times 256^3$ particles with \gadget, and $128^3$ dark matter
particles and a $128^3$ root grid with \enzo. For DM-only runs, the
gravitational softening lengths in these higher resolution \gadget\ runs were
taken to be $1/30$ of the mean dark matter interparticle separation, giving a
dynamic range of $\Le=3840$ and 7680 for $128^3$ and $256^3$ particle runs,
respectively.  For the adiabatic \gadget\ runs, they were taken to be $1/25$
of the mean interparticle separation, giving $\Le=3200$ and 6400 for the
$128^3$ and $256^3$ particle runs, respectively.  All \enzo\ runs used a
maximum refinement ratio of $\Le=4096$.

\begin{figure*}
\epsscale{1.7}
\plotone{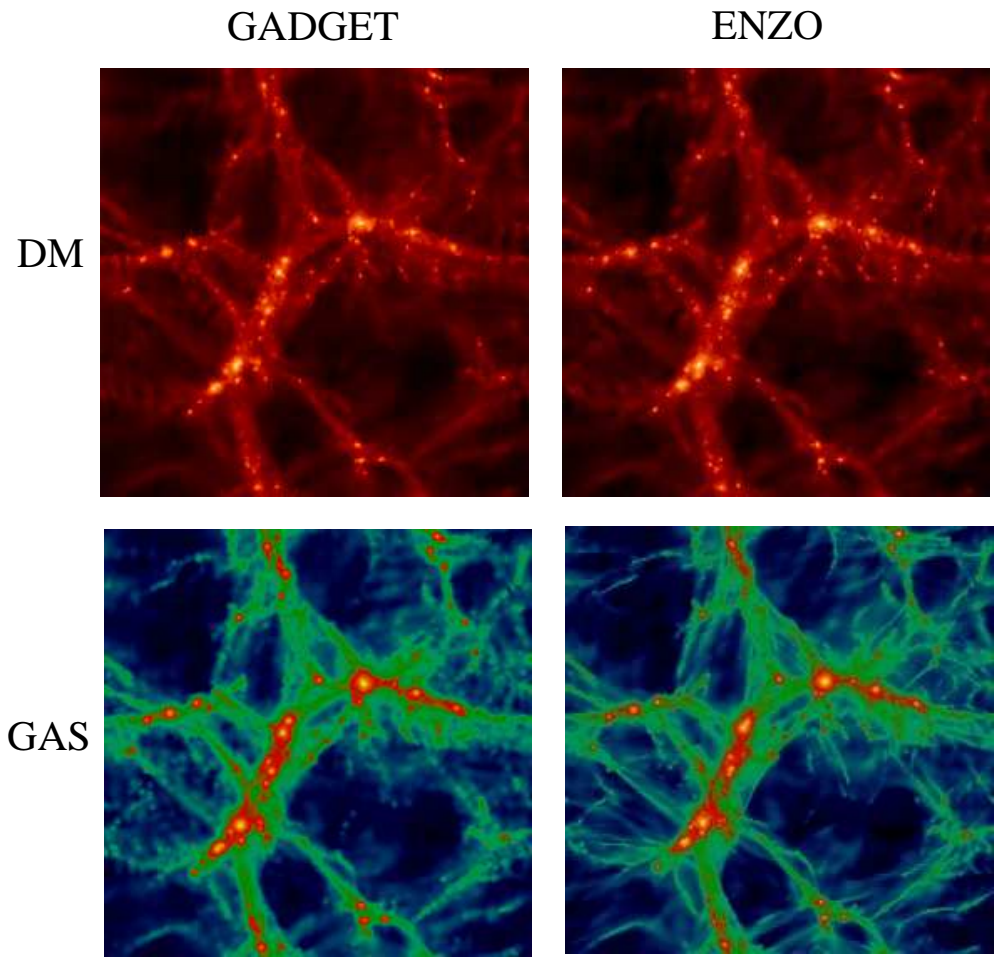}
\caption{Projected dark matter (top row) and gas mass (bottom row) 
distribution for  \gadget\ and \enzo\ in a slab of size 
$3\times 3\times 0.75\, (\himpc)^3$.
For \gadget\ (left column), we used the run with $2\times 64^3$ particles.
For \enzo\ (right column),  the run with $64^3$ dark matter particles 
and $128^3$ root grid was used. 
}
\label{fig.pic}
\end{figure*}

As an example, we show the spatial distribution of the projected dark matter
and gas mass in Figure~\ref{fig.pic} from one of the representative adiabatic gas
runs of \gadget\ and \enzo. The mass distribution in the two simulations are
remarkably similar for both dark matter and gas, except that one can see
slightly finer structures in \gadget\ gas mass distribution compared to that
of \enzo. The good visual agreement in the two runs is very encouraging, and
we will analyze the two simulations quantitatively in the following sections.

\notetoeditor{Put Fig.1 here.}


\section{Simulations with dark matter only}
\label{dm}

According to the currently favored theoretical model of the CDM
theory, the material content of the universe is dominated by as of yet
unidentified elementary particles which interact so weakly that they
can be viewed as a fully collisionless component at spatial scales of
interest for large-scale structure formation. The mean mass density in
this cold dark matter far exceeds that of ordinary baryons, by a
factor of $\sim 5-7$ in the currently favored $\Lambda$CDM
cosmology. Since structure formation in the Universe is primarily
driven by gravity it is of fundamental importance that the
dynamics of the dark matter and the self-gravity of the hydrodynamic
component are simulated accurately by any cosmological code.
In this section we discuss
simulations that only follow dark matter in order to compare
\enzo\ and \gadget\ in this respect.

\subsection{Dark matter power spectrum}
\label{pspec}

\begin{figure*}
\begin{center}
\resizebox{9.0cm}{!}{\includegraphics{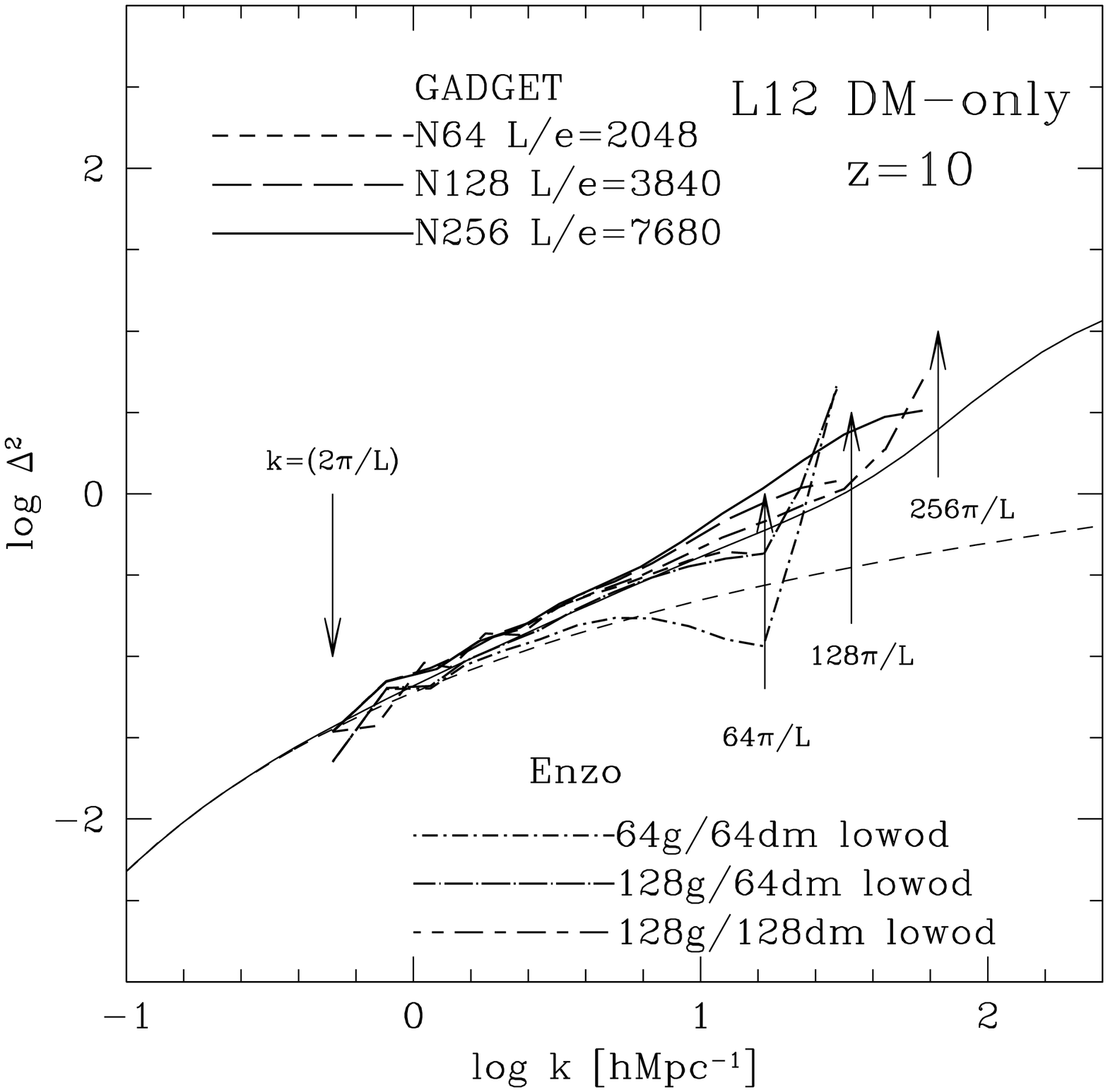}}\\
\vspace{0.1cm}
\resizebox{9.0cm}{!}{\includegraphics{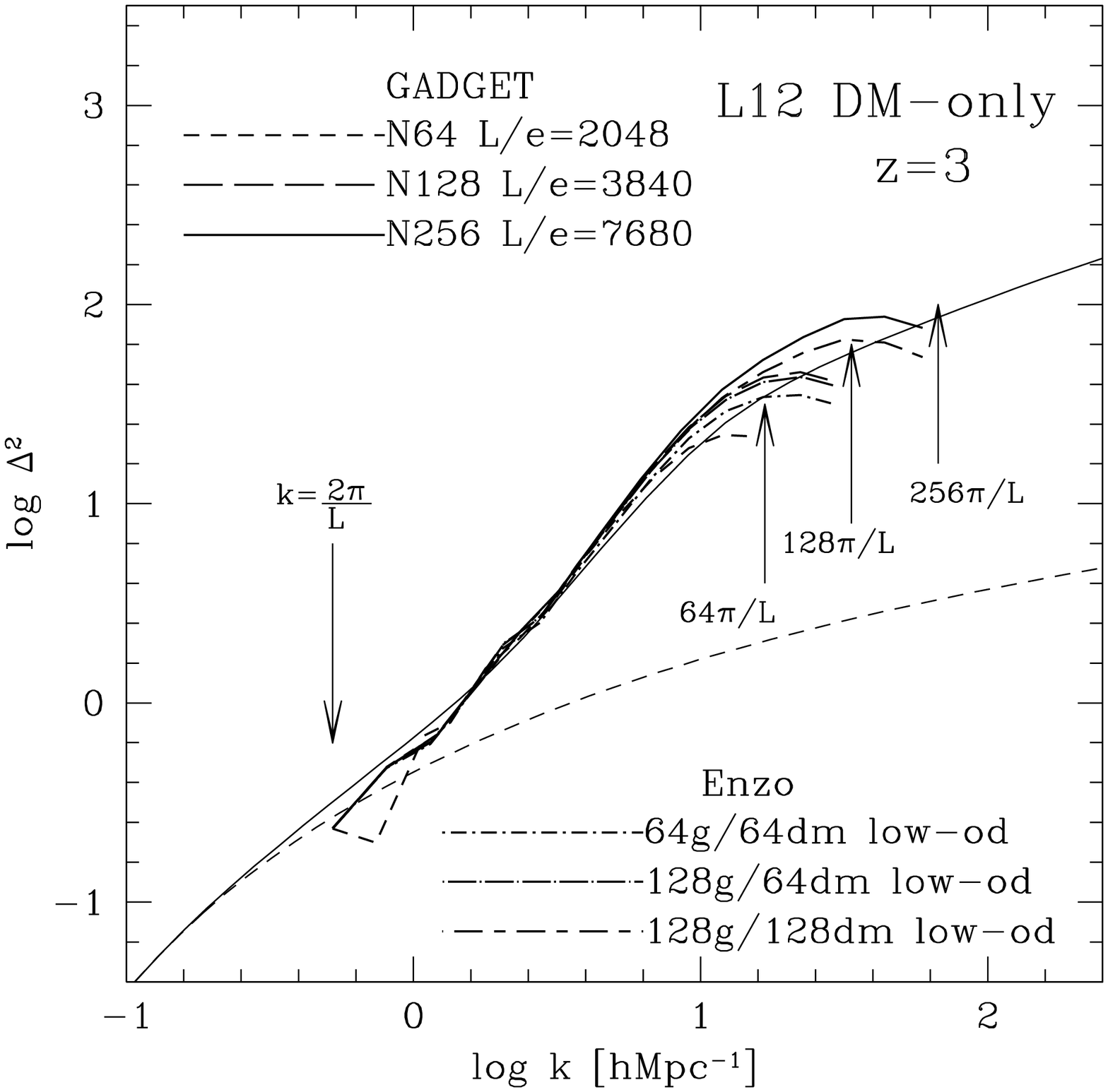}}%
\vspace{0.1cm}
\caption{Dark matter power spectra at $z=10$ and $z=3$ for both \enzo\ 
and \gadget\ simulations with $64^3$ dark matter particles,
 $\Lbox=12~\himpc$ (comoving) and varying spatial resolution.
The short-dashed curve in each panel is the 
linear power spectrum predicted by theory using the transfer function of 
\citet{E-Hu}. The solid curve in each panel is the non-linear power 
spectrum calculated with the \citet{Peacock96} method.
Arrows indicate the largest wavelength that can be accurately represented
in the simulation initial conditions ($k=2\pi/\Lbox$) and those
that correspond to the Nyquist frequencies of $64^3$, $128^3$, and 
$256^3$ \enzo\ root grids.}
\label{fig.pspec}
\end{center}
\end{figure*}

One of the most fundamental quantities to characterize the clustering
of matter is the power spectrum of dark matter density fluctuations.
In Figure~\ref{fig.pspec} we compare the power spectra of DM-only
runs at redshifts $z=10$ and 3. The short-dashed curve is the linearly
evolved power spectrum based on the transfer function of \citet{E-Hu},
while the solid curve gives the expected nonlinear power spectrum
calculated with the \citet{Peacock96} scheme.  We calculate the dark matter
power spectrum in each simulation by creating a uniform grid of dark 
matter densities.  The grid resolution is twice as fine as the mean
interparticle spacing of the simulation (i.e. a simulation with $128^3$ 
particles will use a $256^3$ grid to calculate the power spectrum) and 
densities are generated with the triangular-shaped cloud (TSC) method.  
A fast fourier transform is then performed on the grid of density values
and the power spectrum is calculated by averaging the power in logarithmic 
bins of wavenumber. We do not attempt to correct for shot-noise or the
smoothing effects of the TSC kernel.

The results of all \gadget\ and \enzo\ runs with $128^3$ root grid agree well
with each other at both epochs up to the Nyquist wavenumber. However, the
\enzo\ simulations with a $64^3$ root grid deviate on small scales from the
other results significantly, particularly at $z=10$. This can be understood to
be a consequence of the particle-mesh technique adopted as the gravity solver
in the AMR code, which induces a softening of the gravitational force on the
scale of one mesh cell (this is a property of all PM codes, not just \enzo).
To obtain reasonably accurate forces down to the scale of the interparticle
spacing, at least two cells per particle spacing are therefore required at the
outset of the calculation.  In particular, the force accuracy of \enzo\ is
much less accurate at small scales at early times when compared to \gadget\ 
because before significant overdensities develop the code does not adaptively
refine any regions of space (and therefore increased force resolution to
include small-scale force corrections).  \gadget\ is a tree-PM code -- at
short range, forces on particles are calculated using the tree method, which
offers a force accuracy that is essentially independent of the clustering
state of the matter down to the adopted gravitational softening length
(see Section~\ref{gravmethod} for details).

However, as the simulation progresses in time and dark matter begins
to cluster into halos, the force calculation by \enzo\ becomes more
accurate as additional levels of grids are adaptively added to the
high density regions, reducing the discrepancy seen between \enzo\ and
\gadget\ at redshift $z=10$ to something much smaller at $z=3$.

\notetoeditor{Put Fig.2 here.}

\notetoeditor{Put Fig.3 here.}

\notetoeditor{Put Fig.4 here.}


\subsection{Halo dark matter mass function and halo positions}
\label{dmhalo}

We have identified dark matter halos in the simulations using a
standard friends-of-friends algorithm with a linking length of 0.2 in
units of the mean interparticle separation. In this section, we
consider only halos with more than 32 particles.  
We obtained nearly identical results to those described
in this section using the HOP halo finder \citep{EisHut98}.

\begin{figure*}
\epsscale{1.7}
\plotone{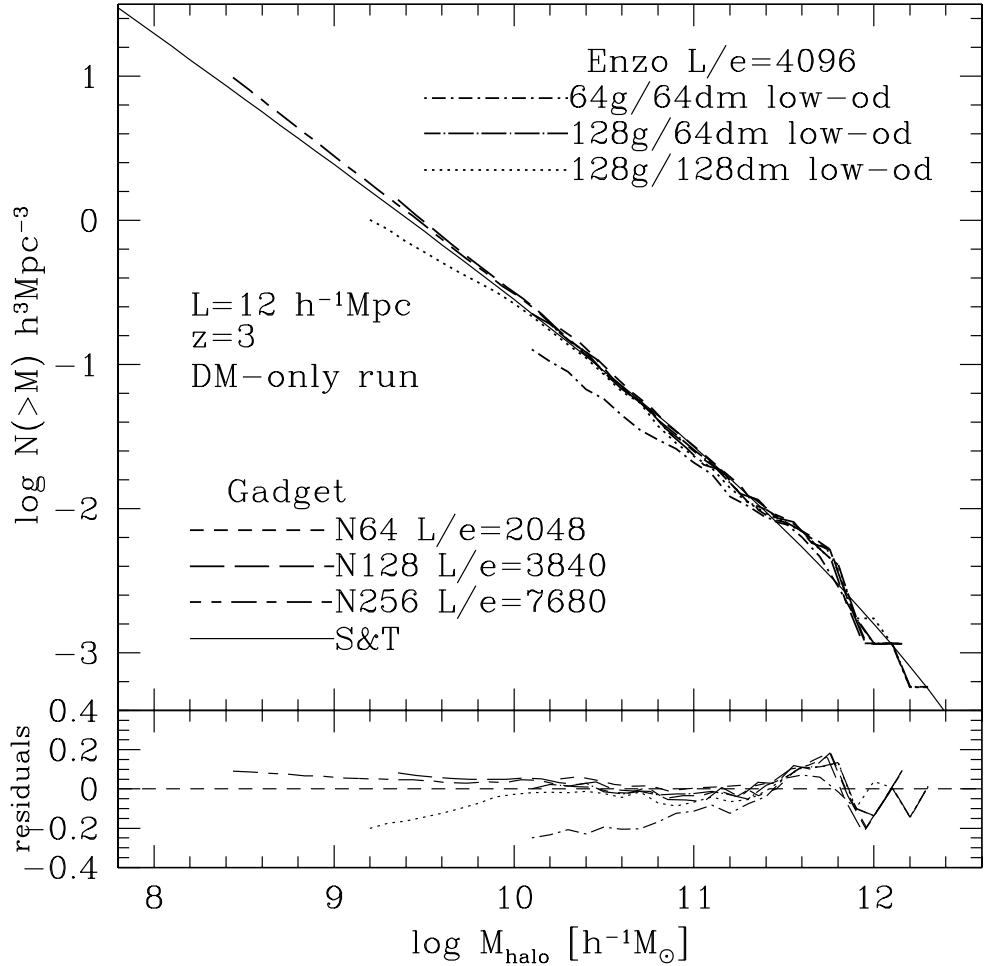}
\caption{Cumulative mass functions at $z=3$ for dark matter-only \enzo\ \& 
\gadget\ runs with $64^3$ particles and a comoving boxsize of $\Lbox=~12\himpc$.  
All \enzo\ runs have $\Le=4096$.
The solid black line denotes the \citet{Sheth99} mass function.
In the bottom panel, we show the residual in logarithmic space
with respect to the Sheth-Tormen mass function, i.e., 
$\log$(N$>$M)$-\log$(S\&T). 
}
\label{fig.dmmf}
\end{figure*}

In Figure~\ref{fig.dmmf}, we compare the cumulative DM halo mass function
for several simulations with $64^3$, $128^3$ and $256^3$ 
dark matter particles as a
function of $\Le$ and particle mass.  In the bottom panel, we show the residual in 
logarithmic space with respect to the Sheth-Tormen mass function, i.e., 
$\log$(N$>$M)$-\log$(S\&T). 
The agreement between \enzo\ and \gadget\
simulations at the high-mass end of the mass function is reasonable,
but at lower masses there is a systematic difference between the two
codes. The \enzo\ run with $64^3$ root grid contains significantly
fewer low mass halos compared to the \gadget\ simulations. Increasing
the root grid size to $128^3$ brings the low-mass end of the \enzo\
result closer to that of \gadget.

This highlights the importance of the size of the root grid in
the adaptive particle-mesh method based AMR simulations. Eulerian simulations
using the particle-mesh technique require a root grid twice
as fine as the mean interparticle separation in order to achieve a
force resolution at early times 
comparable to tree methods or so-called P$^3$M
methods \citep{Efstathiou85}, which supplement the softened
PM force with a direct particle-particle (PP) summation on the scale
of the mesh.  Having a conservative refinement criterion together with
a coarse root grid in AMR is not sufficient to improve the low mass
end of the halo mass function because the lack of force resolution at
early times effectively results in a loss of small-scale power, which
then prevents many low mass halos from forming.

We have also directly compared the positions of individual dark matter
halos identified in a simulation with the same initial conditions, run
both with \gadget\ and \enzo. This run had $64^3$ dark matter
particles and a $\Lbox=12\himpc$ box size. For \gadget, we used a
gravitational softening equivalent to $\Le=2048$. For \enzo, we used a
$128^3$ root grid, a low overdensity threshold for the refinement
criteria, and we limited refinements to a dynamic range of $\Le=4096$ (5 
total levels of refinement).

In order to match up halos, we apply the following method to identify
``pairs'' of halos with approximately the same mass and center-of-mass
position. First, we sort the halos in order of decreasing mass, and
then select a halo from the massive end of one of the two
simulations (i.e. the beginning of the list). 
Starting again from the massive end, we then search the
other list of halos for a halo within a distance of $\rmax = f_R
\Delta$, where $\Delta$ is the mean interparticle separation ($1/64$
of the boxsize in this case) and $f_{R}$ is a dimensionless number
(chosen here to be either 0.5 or 1.0).  If the halo masses are also
within a fraction $f_{M}$ of one another, then the two halos in
question are counted as a `matched pair' and removed from the lists to
avoid double-counting. This procedure is continued until there are no
more halos left that satisfy these criteria.

\begin{figure*}
\epsscale{1.7}
\plotone{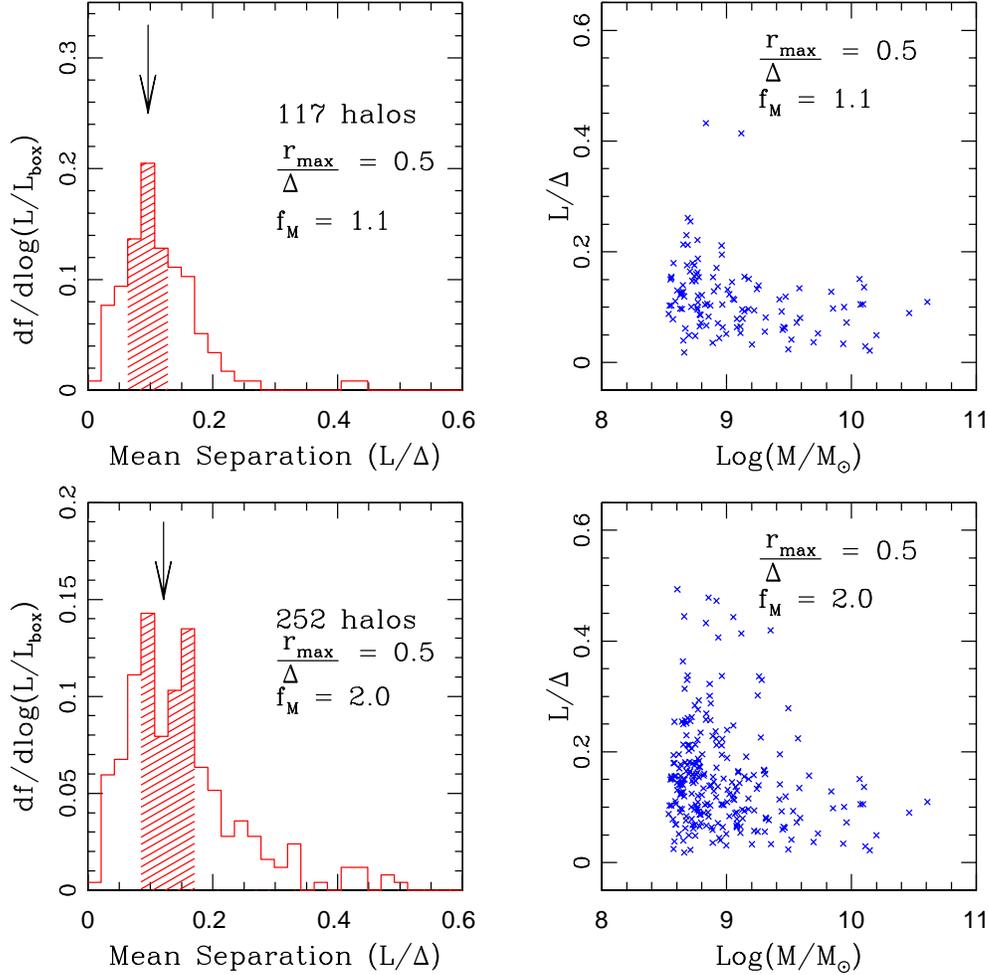}
\caption{{\it Left column}:  Probability distribution function of 
the number of dark matter halos as a function of the separation 
of the matched halo pair in corresponding \enzo\ and \gadget\ 
simulations (see text for the details of the runs used in this 
comparison). The separation is in units of the initial mean 
interparticle separation, $\Delta$ .  The shaded region in the 
distribution function shows the quartiles on both sides of the median 
value (which is shown by the arrows) of the distribution.
{\it Right column}:  Separation of each pair (in units of $\Delta$)
vs. mean halo mass of each pair.  The top row is of pairs whose 
masses agree to within 10\% (i.e. $f_M=1.1$) and the bottom row 
is of pairs whose masses agree to within a factor of two 
(i.e. $f_M=2.0$).}
\label{fig.dmhalo}
\end{figure*}

In the left column of Figure~\ref{fig.dmhalo}, we show the
distribution of pair separations obtained in this way. The arrow
indicates the median value of the distribution, and the quartile on
each side of the median value is indicated by the shaded region. The
values of $\rmax$ and $f_M$ are also shown in each panel. A
conservative matching-criterion that allows only a 10\% deviation in
halo mass and half a cell of variation in the position
(i.e. $\rmax=0.5 \Delta$, $f_M= 1.1$) finds only 117 halo pairs (out
of $\sim 292$ halos in each simulation) with a median separation of
$0.096 \Delta$ between the center-of-mass positions of halos.
Increasing $\rmax$ to $1.0~\Delta$ does very little to increase
the number of matched halos. Keeping $\rmax = 0.5\Delta$ and
increasing $f_{M}$ to 2.0 gives us 252 halo pairs with a median
separation of $0.128\Delta$.  Increasing $f_{M}$ any further does
little to increase the number of matched pairs, and looking further
away than $\rmax = 1.0 \Delta$ produces spurious results in some
cases, particularly for low halo masses.

This result therefore suggests that the halos are typically in almost the same
places in both simulations, but that their individual masses show somewhat
larger fluctuations. Note however that a large fraction of this scatter simply
stems from noise inherent in the group sizes obtained with the halo finding
algorithms used.  The friends-of-friends algorithm often links (or not yet
links) infalling satellites across feeble particle bridges with the halo, so
that the numbers of particles linked to a halo can show large variations
between simulations even though the halo's virial mass is nearly identical in
the runs.  We also tested the group finder HOP \citep{EisHut98}, but found
that it also shows significant noise in the estimation of halo masses.  It may
be possible to reduce the latter by experimenting with the adjustable
parameters of this group finder (one of which controls the ``bridging
problem'' that the friends-of-friends method is susceptible to), but we have
not tried this.

In the right panels of Figure~\ref{fig.dmhalo}, we plot the separation
of halo pairs against the average mass of the two halos in question.
Clearly, pairs of massive halos tend to have smaller separations than
low mass halos.  Note that some of the low mass halos with large
separation ($L/\Del > 0.4$) could be false identifications.  It is
very encouraging, however, that the massive halos in the two simulations
generally lie within $1/10$ of the initial mean interparticle
separation.  The slight differences in halo positions may be caused by 
timing differences between the two simulation codes.  


\subsection{Halo dark matter substructure}
\label{dmsub}

Another way to compare the solution accuracy of the N-body problem in the two
codes is to examine the substructure of dark matter halos.  The most massive
halos in the $128^3$ particle dark matter-only simulations discussed in this
paper have approximately 11,000 particles, which is enough to marginally
resolve substructure. We look for gravitationally-bound substructure using the
{\small SUBFIND} method described in \citet{Springel01}, which we briefly
summarize here for clarity.  The process is as follows: a Friends-of-Friends
group finder is used (with the standard linking length of 0.2 times the mean
interparticle spacing) to find all of the dark matter halos in the
simulations.  We then select the two most massive halos in the calculation
(each of which has at least 11,000 particles in both simulations) and analyze
them with the subhalo finding algorithm. This algorithm first computes a local
estimate of the density at the positions of all particles in the input group,
and then finds locally overdense regions using a topological method. Each of
the substructure candidates identified in this way is then subjected to a
gravitational unbinding procedure where only particles bound to the
substructure are kept.  If the remaining self-bound particle group has more
than some minimum number of particles it is considered to be a subhalo.  We
use identical parameters for the Friends-of-Friends and subhalo detection
calculations for both the \enzo\ and \gadget\ dark matter-only calculations.

\begin{figure*}
\epsscale{1.4}
\plotone{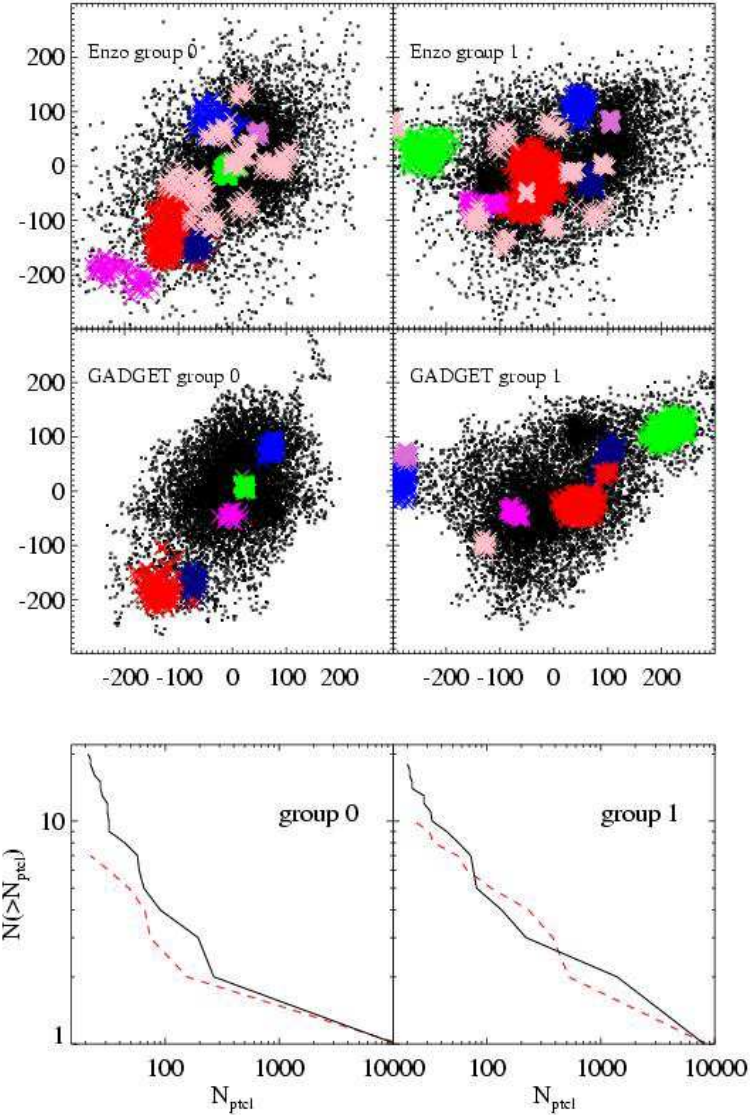}
\caption{Dark matter substructure in both \enzo\ and \gadget\ 
  dark matter-only calculations with $128^3$ particles.  The \enzo\ 
  simulations use the ``low overdensity'' refinement parameters.  Left column:
  data from the most massive halo in the simulation volume.  Right column:
  second most massive halo.  Top row: Projected dark matter density for halos
  in the \enzo\ simulation with substructure color-coded.  Middle row:
  projected dark matter density for \gadget\ simulations.  Bottom row: Halo
  substructure mass function for each halo with both \enzo\ and \gadget\ 
  results plotted together, with units of number of halos greater than a given
  mass on the y axis and number of particles on the x axis.  In these
  simulations the dark matter particle mass is $9.82 \times 10^7~M_{\odot}$,
  resulting in total halo masses of $\sim 10^{12}~M_{\odot}$.  In the top and
  middle rows subhalos with the same color correspond to the most massive,
  second most massive, etc. subhalos.  In the \enzo\ calculation all subhalos
  beyond the 10th most massive are shown using the same color.  Both sets of
  halos have masses of $\sim 10^{12}~M_{\odot}$ The x and y axes in the top
  two rows are in units of comoving kpc/h.  In the bottom row, \enzo\ results
  are shown as a black solid line and \gadget\ results are shown as a red
  dashed line.}
\label{fig.dmsub}
\end{figure*}

Figure~\ref{fig.dmsub} shows the projected dark matter density distribution
and substructure mass function for the two most massive halos in the $128^3$
particle DM-only calculations for both \enzo\ and \gadget, which have dark
matter masses close to $M_{\rm halo} \sim 10^{12} M_{\odot}$.  Bound subhalos
are indicated by different colors, with identical colors being used in both
simulations to denote the most massive subhalo, second most massive, etc.
Qualitatively, the halos have similar overall morphologies in both 
calculations, though there are some differences in the substructures. 
The masses of these two parent halos in the \enzo\ calculation are 
$8.19 \times 10^{11}~M_{\odot}$ and $7.14 \times 10^{11}~M_{\odot}$, and we 
identify total 20 and 18 subhalos, respectively.  The corresponding halos 
in the \gadget\ calculation have masses of $8.27 \times 10^{11}~M_{\odot}$ 
and $7.29 \times 10^{11}~M_{\odot}$, and they have 7 and 10 subhalos. Despite 
the difficulty of \enzo\ in fully resolving the low-mass end of the halo 
mass function, the code apparently has no problem in following dark matter 
substructure within large halos, and hosts larger number of small subhalos 
than the \gadget\ calculation. Some corresponding subhalos in the two 
calculations appear to be slightly off-set. Overall, the agreement of the 
substructure mass functions for the intermediate mass regime of subhalos 
is relatively good and within the expected noise.

It is not fully clear what causes the observed differences in halo 
substructure between the two codes. It may be due to lack of spatial and/or 
dark matter particle mass resolution in the calculations -- typically 
simulations used for substructure studies have at least an order of 
magnitude more dark matter particles per halo than we have here.
It is also possible that systematics in the grouping algorithm are 
responsible for some of the differences.

\notetoeditor{Put Fig.5 here.}


\section{Adiabatic simulations}
\label{baryon}

In this section, we start our comparison of the fundamentally
different hydrodynamical algorithms of \enzo\ and \gadget.  It is
important to keep in mind that a direct comparison between the
AMR and SPH methods when applied to cosmic structure formation will
always be convolved with a comparison of the gravity solvers of the
codes. This is because the process of structure formation is primarily
driven by gravity, to the extent that hydrodynamical forces are
subdominant in most of the volume of the universe. Differences that
originate in the gravitational dynamics will in general induce
differences in the hydrodynamical sector as well, and it may not
always be straightforward to cleanly separate those from genuine
differences between the AMR and SPH methods themselves.  Given that
the dark matter comparisons indicate that one must be careful to 
appropriately resolve dark matter forces at early times unless 
relatively fine root grids are used for \enzo\ calculations, 
it is clear that any difference found between the codes
needs to be regarded with caution until confirmed with AMR
simulations of high gravitational force resolution.

Having made these cautionary remarks, we will begin our comparison
with a seemingly trivial test of a freely expanding universe without
perturbations, which is useful to check conservation of
entropy (for example). After that, we will compare the gas properties
found in cosmological simulations of the $\Lambda$CDM model in more
detail.

\subsection{Unperturbed adiabatic expansion test}
\label{unpert}

\notetoeditor{Put Fig.6 here.}

Unperturbed ideal gas in an expanding universe should follow Poisson's
law of adiabatic expansion: $T\propto V^{\gamma-1} \propto
\rho^{1-\gamma}$.  Therefore, if we define entropy as $S\equiv
T/\rho^{\gamma-1}$, it should be constant for an adiabatically
expanding gas. 

This simple relation suggests a straightforward test of how well the
hydrodynamic codes described in Section~\ref{code} conserve
entropy \citep{He93}. To this end, we set up unperturbed simulations for both
\enzo\ and \gadget\ with $16^3$ grid cells or particles,
respectively. The runs are initialized at $z=99$ with uniform density
and temperature $T = 10^4 \K$.  This initial temperature was
deliberately set to a higher value than expected for the real universe
in order to avoid hitting the temperature floor set in the codes while
following the adiabatic cooling of gas due to the expansion of the
universe.  The box was then allowed to expand
until $z=3$.  \enzo\ runs were performed using both the PPM and \zeus\
algorithms and \gadget\ runs were done with both `conventional' and
the `entropy conserving' formulation of SPH.

\begin{figure*}
\epsscale{1.8}
\plotone{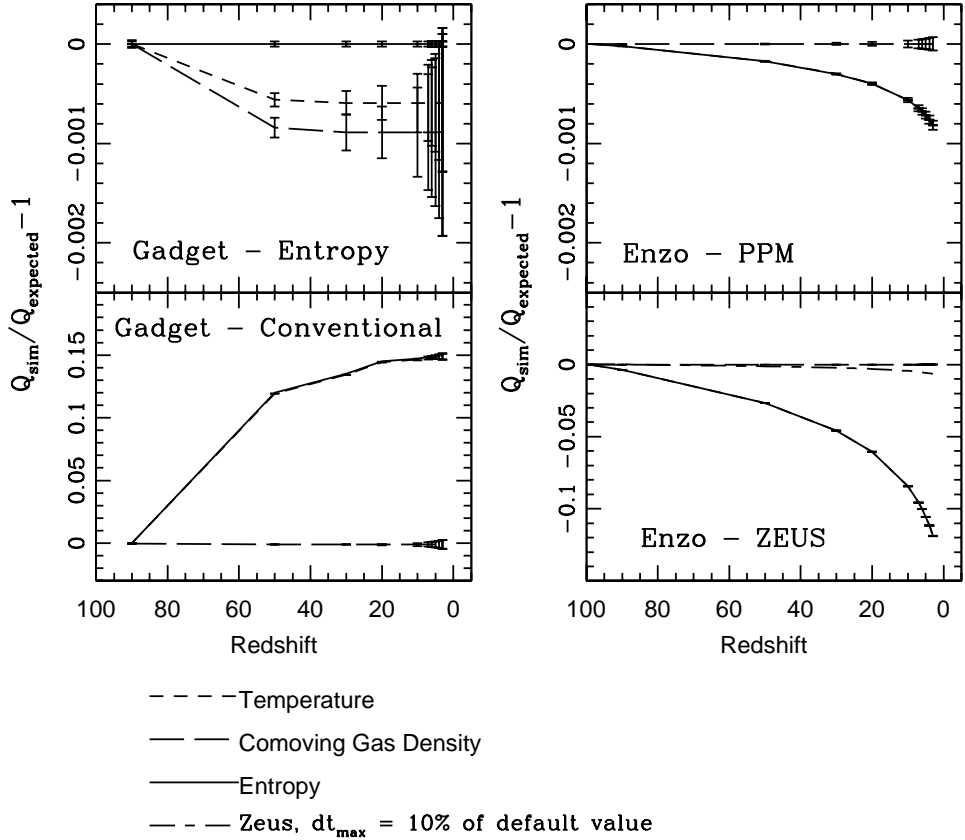}
\caption{Fractional deviations from the expected adiabatic relation 
for temperature, comoving gas density and entropy as a function of redshift 
in simulations of unperturbed adiabatic expansion test.
{\it Left column:}  the `entropy conserving' formulation of SPH (top panel)
and the `conventional' formulation (bottom panel).  {\it Right column:}  
The PPM (top panel) and \zeus\ (bottom panel) hydrodynamic methods 
in \enzo.  Error bars in all panels show the variance of each quantity.
The short-long-dashed line in the bottom right panel shows the case where 
the maximum timestep is limited to be 1/10 of the default maximum.  Note the
difference in scales of the y axes in the bottom row.}
\label{fig.unpert}
\end{figure*}

In Figure~\ref{fig.unpert} we show the fractional deviation from the
expected adiabatic relation for density, temperature, and entropy.
The \gadget\ results (left column) show that the `entropy conserving'
formulation of SPH preserves the entropy very well, as expected.
There is a small net decrease in temperature and density of only $\sim
0.1$\%, reflecting the error of SPH in estimating the mean density.
In contrast, in the `conventional' SPH formulation the temperature and
entropy deviate from the adiabatic relation by 15\%, while the
comoving density of each gas particle remains constant. This systematic
drift is here caused by a small error in estimating the local velocity
dispersion owing to the expansion of the universe. In physical
coordinates, one expects $\vec{\nabla}\cdot \vec{v} = 3 H(a)$, but in
conventional SPH, the velocity divergence needs to be estimated with a
small number of discrete particles, which in general will give a
result that slightly deviates from the continuum expectation of $3
H(a)$. In our test, this error is the same for all particles, without
having a chance to average out for particles with different neighbor
configurations, hence resulting in a substantial systematic drift. In
the entropy formulation of SPH, this problem is absent by
construction.

In the \enzo/\ppm\ run (top right panel), there is a net
decrease of only $\sim 0.1$\% in temperature and entropy, whereas in
\enzo/\zeus\ (bottom right panel), the temperature and entropy drop by
12\% between $z=99$ and $z=3$.  The comoving gas density remains
constant in all \enzo\ runs.  In the bottom right panel, the
short-long-dashed line shows an \enzo/\zeus\ run where we lowered the
maximum expansion  of the simulation volume during a single
timestep (i.e. $\Delta a/a$, where $a$ is the scale factor) by a
factor of 10. This results in a factor of $\sim 10$ reduction of the
error, such that the fractional deviation from the adiabatic relation
is only about 1\%.  This behavior is to be expected since the \zeus\
hydrodynamic algorithm is formally first-order-accurate in time in an
expanding universe.

In summary, these results show that both the \enzo/\zeus\ hydrodynamic
algorithm and the conventional SPH formulation in \gadget\ have
problems in reliably conserving entropy. However, these problems are
essentially absent in \enzo/\ppm\ and the new SPH formulation
of \gadget.


\subsection{Differential distribution functions of gas properties}
\label{diff-dist}

\begin{figure*}
\epsscale{2.0}
\plotone{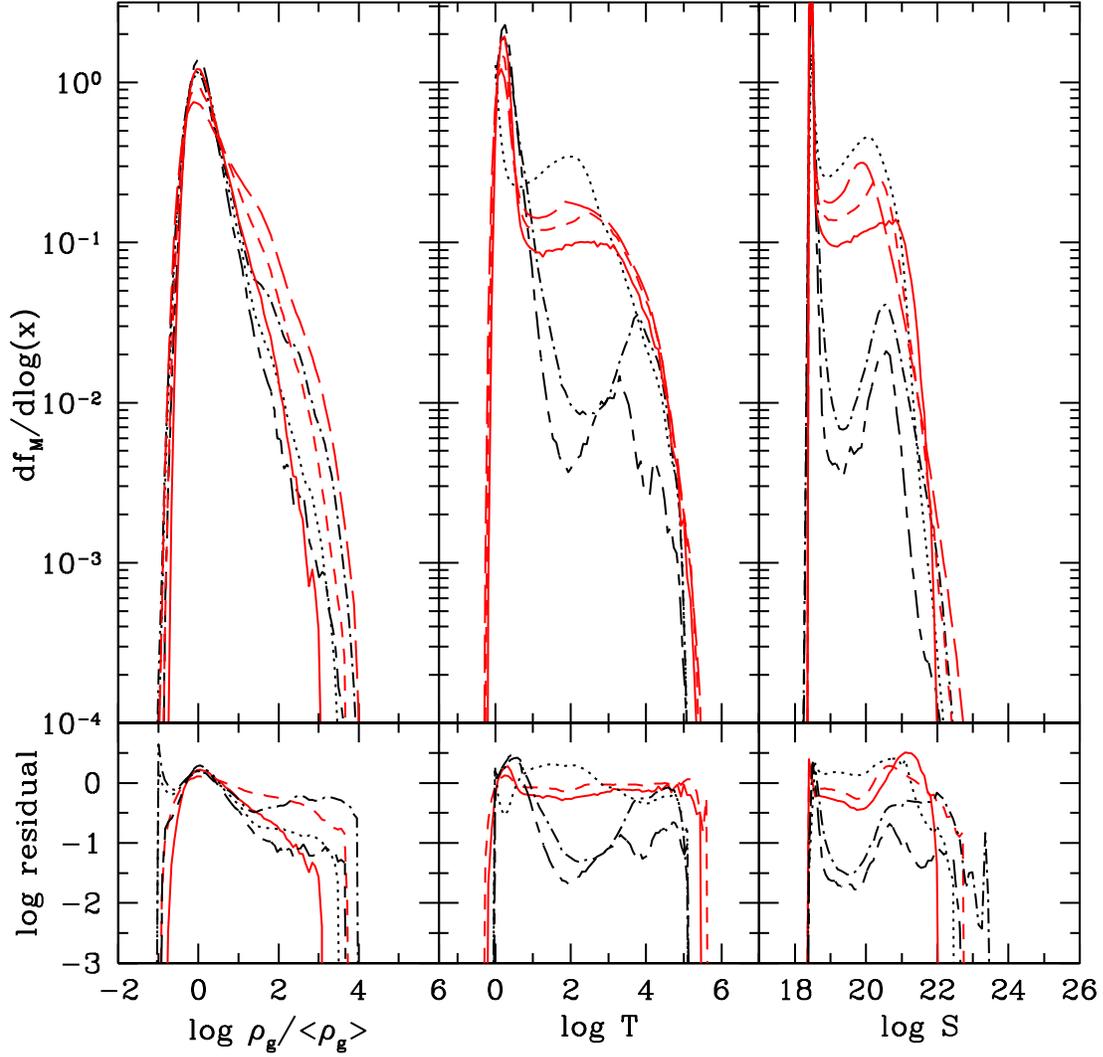}
\caption{Probability distribution functions of gas mass as functions
of gas overdensity (left column), temperature (middle column) and 
entropy (right column) at $z=10$.
For \gadget, runs with $2\times 64^3$ ({\it red solid line}), $2\times 128^3$ 
({\it red short-dashed line}) and $2\times 256^3$ ({\it red long-dashed line}) 
particles are shown.  The dynamic range of the \enzo\ simulations were 
fixed to $\Le = 4096$, but the particle numbers and the root grid size were 
varied between $64^3$ and $128^3$.  Both the \zeus\ and \ppm\ hydro 
methods were used in the \enzo\ calculations. The \enzo\ line types are:
128g/128dm \ppm\ lowod ({\it black dash-dotted line}), 128g/128dm \zeus\ 
({\it black dotted line}), and 64g/64dm \ppm\ lowod ({\it black long 
dash-short dashed line}). In the bottom panels, the residuals in logarithmic 
scale with respect to the \gadget\ N256 run are shown.}
\label{fig.diffdf-1}
\end{figure*}

\begin{figure*}
\epsscale{2.0}
\plotone{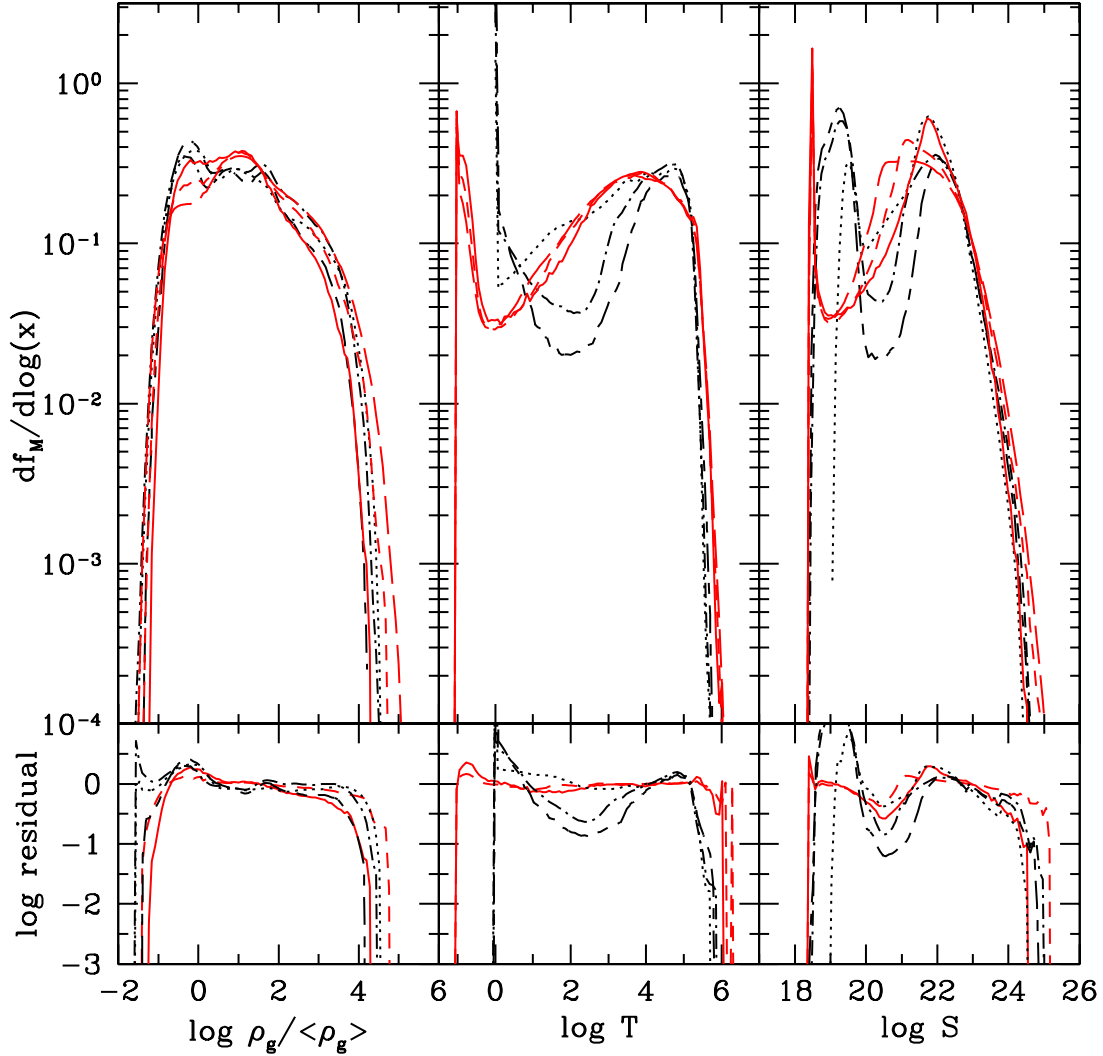}
\caption{Probability distribution functions of gas mass as functions
of gas overdensity (left column), temperature (middle column) and 
entropy (right column) at $z=3$.
For \gadget, runs with $2\times 64^3$, $2\times 128^3$ and $2\times 256^3$ 
particles were used.  The dynamic range of the \enzo\ simulations were 
fixed to $\Le = 4096$, but the particle numbers and the root grid size were 
varied between $64^3$ and $128^3$ (e.g. 64dm/128grid means $64^3$ DM 
particles and $128^3$ root grid).  Both the \zeus\ and \ppm\ hydro 
methods were used in the \enzo\ calculations, as shown in the figure 
key.  Lines are identical to those in Figure~\ref{fig.diffdf-1}.
In the bottom panels, we show the residuals in logarithmic scale with 
respect to the \gadget\ N256 run.}
\label{fig.diffdf-2}
\end{figure*}

We now begin our analysis of gas properties found in full cosmological
simulations of structure formation.  In Figures~\ref{fig.diffdf-1} 
and~\ref{fig.diffdf-2} we
show mass-weighted one-dimensional differential probability
distribution functions of gas density, temperature and entropy, for
redshifts $z=10$ (Figure~\ref{fig.diffdf-1}) and $z=3$ 
(Figure~\ref{fig.diffdf-2}). We 
compare results for \gadget\ and \enzo\ simulations at different numerical
resolution, and run with both the \zeus\ and \ppm\
formulations of \enzo.

At $z=10$, effects owing to an increase of resolution are clearly seen
in the distribution of gas overdensity (left column), with runs of
higher resolution reaching higher densities earlier than those of
lower resolution. However, this discrepancy becomes smaller at $z=3$
because lower resolution runs tend to `catch up' at late times,
indicating that then more massive structures, which are also resolved
in the lower resolution simulations, become ever more important.  One
can also see that the density distribution becomes wider at $z=3$
compared to those at $z=10$, reaching to higher gas densities at lower
redshift.

At $z=3$, both \enzo\ and \gadget\ simulations agree very well at
$\log T > 3.5$ and $\log S > 21.5$, with a characteristic shoulder in
the temperature (middle column) and a peak in the entropy (right
column) distributions at these values.  This can be understood with a
simple analytic estimate of gas properties in dark matter halos.  We
estimate the virial temperature of a dark matter halo with mass
$10^8\Msun$ ($10^{11}\Msun$) at $z=3$ to be $\log T=3.7$ ($5.7$).
Assuming a gas overdensity of 200, the corresponding entropy is $\log
S = 21.9$ (23.9).  The good agreement in the distribution functions at
$\log T > 3.5$ and $\log S > 21.5$ therefore suggests that the
properties of gas inside the dark matter halos agree reasonably well
in both simulations. The gas in the upper end of the
distribution is in the most massive halos in the simulation, with
masses of $\sim 10^{11}\Msun$ at $z=3$.  \enzo\ has a built-in
temperature floor of $1 K$, resulting in an artificial feature
 in the temperature and entropy profiles at $z=3$.  \gadget\ also has a
temperature floor, but it is set to $0.1 K$ and is much less noticeable 
since that temperature is not attained in this simulation.  Note that the 
entropy floor stays at the constant value of $\log S_{\rm init}=18.44$ 
for all simulations at both redshifts.

However, there are also some interesting differences in the
distribution of temperature and entropy between \enzo/{\small PPM} and
the other methods for gas of low overdensity. \enzo/{\small PPM}
exhibits a `dip' at intermediate temperature ($\log T\sim 2.0$) and
entropy ($\log S\sim 20$), whereas \enzo/\zeus\ and \gadget\ do not
show the resulting bimodal character of the distribution. We will
revisit this feature when we examine two dimensional phase-space
distributions of the gas in Section~\ref{2d-dist-S}, and again in 
Section~\ref{viscosity} when we examine numerical effects due to
artificial viscosity.  In general, the
\gadget\ results appear to lie in between those obtained with
\enzo/\zeus\ and \enzo/{\small PPM}, and are qualitatively more
similar to the \enzo/\zeus\ results.

\notetoeditor{Put Fig.7 here.}

\notetoeditor{Put Fig.8 here.}


\subsection{Cumulative distribution functions of gas properties}
\label{cum-dist}

\begin{figure*}
\epsscale{2.0}
\plotone{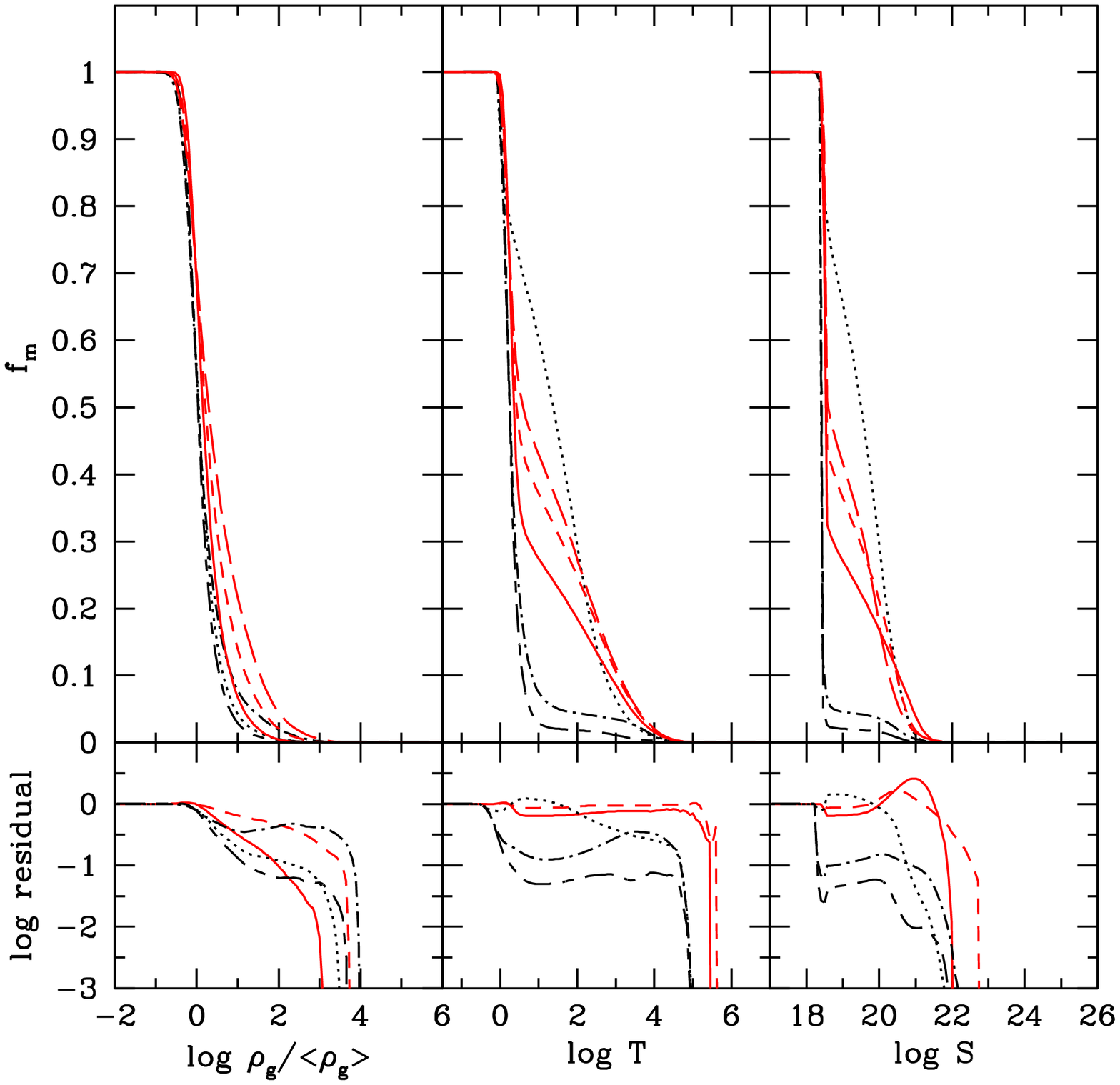}
\caption{Cumulative distribution functions of gas mass as functions
of comoving gas overdensity (left column), temperature (middle column) and 
entropy (right column) at $z=10$.
The simulations and the line types used here are the same as in 
Figures~\ref{fig.diffdf-1} and~\ref{fig.diffdf-2}.
In the bottom panels, we show the residuals in logarithmic scale with 
respect to the \gadget\ N256 run.}
\label{fig.cumdf-1}
\end{figure*}

\begin{figure*}
\epsscale{2.0}
\plotone{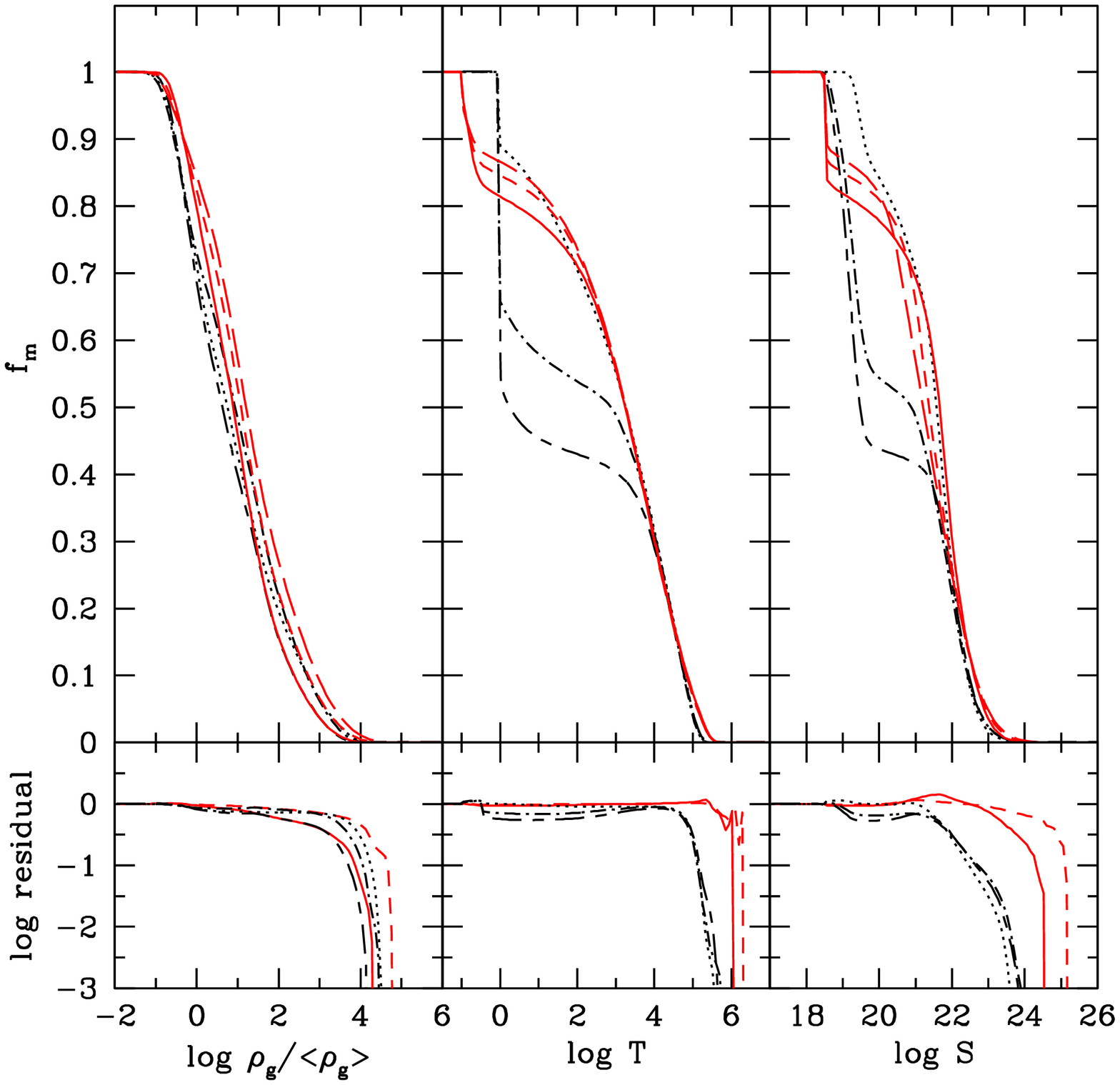}
\caption{Cumulative distribution functions of gas mass as functions
of comoving gas overdensity (left column), temperature (middle column) and 
entropy (right column) at $z=3$.
The simulations and the line types used here are the same as in 
Figures~\ref{fig.diffdf-1} and~\ref{fig.diffdf-2}.
In the bottom panels, we show the residuals in logarithmic scale with 
respect to the \gadget\ N256 run.}
\label{fig.cumdf-2}
\end{figure*}

In this section we study cumulative distribution functions of the quantities
considered above, highlighting the quantitative differences in the
distributions in a more easily accessible way.  In 
Figures~\ref{fig.cumdf-1} and~\ref{fig.cumdf-2} we show the mass-weighted 
cumulative distribution functions of gas overdensity, temperature and 
entropy at $z=10$ (Figure~\ref{fig.cumdf-1}) and $z=3$ 
(Figure~\ref{fig.cumdf-2}). The measurements parallel those described in 
Section~\ref{diff-dist}, and were done for the same simulations.

We observe similar trends as before.  At $z=10$ in the \gadget\
simulations, 70\% of the total gas mass is in regions above the mean
density of baryons, but in \enzo, only 50\% is in such regions. This
mass fraction increases to 80\% in \gadget\ runs, and to 70\% in
\enzo\ runs at $z=3$, as more gas falls into the potential wells of
dark matter halos.

More distinct differences can be observed in the distribution of
temperature and entropy. At $z=10$, only $10-20\%$ of the total gas
mass is heated to temperatures above $\log T = 0.5$ in \enzo/\ppm, 
whereas this fraction is $70-75\%$ in \enzo/\zeus, and $35-55\%$
in \gadget. At $z=3$, the mass fraction that has temperature $\log T>
0.5$ is $40-60\%$ for \enzo/\ppm, and $\sim 80\%$ for both
\enzo/\zeus\ and \gadget.  Similar mass fractions can be observed
for gas with entropy $\log S>18.5-19.0$.

In summary, these results show that both \gadget\ and particularly
\enzo/\zeus\ tend to heat up a significant amount of gas at earlier
times than \enzo/\ppm. This may be related to differences in
the parameterization of numerical viscosity, a topic that we will
discuss in more detail in Section~\ref{viscosity}.

\notetoeditor{Put Fig.9 here.}

\notetoeditor{Put Fig.10 here.}


\subsection{Phase diagrams}
\label{2d-dist-S}

\notetoeditor{Put Fig.11 here.}

\begin{figure*}
\epsscale{2.2}
\plotone{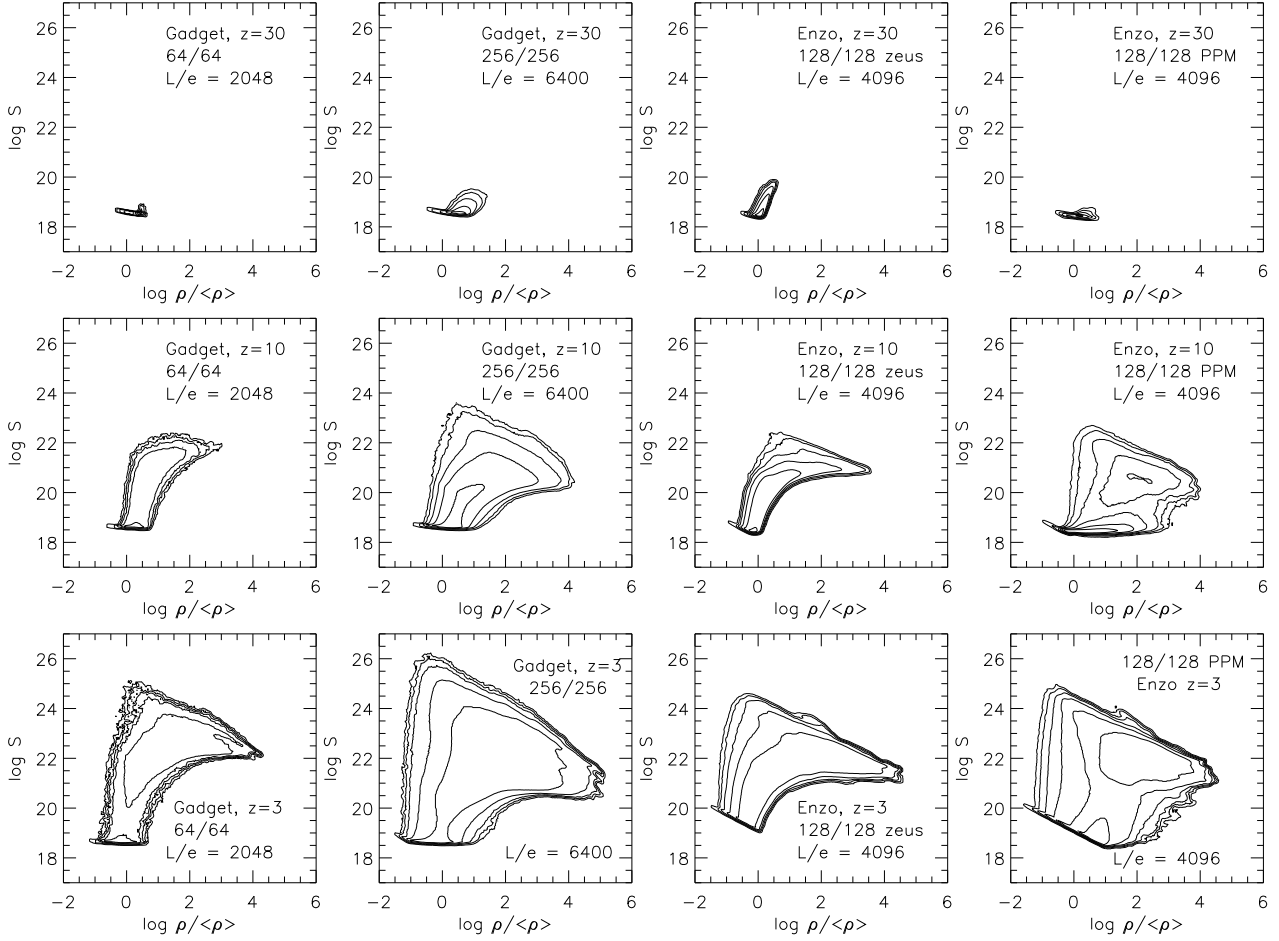}
\caption{
Redshift evolution of the two dimensional mass-weighted distribution 
of gas entropy  vs. gas overdensity for four representative \enzo\
and \gadget\ simulations. Rows correspond to (from top to bottom) $z=30$, 10 
and 3. In each panel six contours are evenly spaced from 0 to the 
maximum value in logarithmic scale, with the scale being identical in all 
simulations at a given redshift to allow for direct comparison.
{\it Column 1}: \gadget, $2\times 64^3$ particles, $\Le = 2048$.  
{\it Column 2}: \gadget, $2\times 256^3$ particles, $\Le = 6400$.  
{\it Column 3}:  \enzo/\zeus\ hydro, $128^3$ DM particles, $128^3$ root grid, 
$\Le = 4096$.  Column 4: \enzo/{\small PPM} hydro, $128^3$ DM particles, 
$128^3$ root grid, $\Le = 4096$. The increasing minimum entropy with
decreasing overdensity in the Enzo results is an artifact of imposing
a temperature floor--a numerical convenience.}
\label{fig.2d_S_evol}
\end{figure*}

In Figure~\ref{fig.2d_S_evol} we show the redshift evolution of the
mass-weighted two-dimensional distribution of entropy vs. gas
overdensity for redshifts $z=30$, 10 and 3 (top to bottom rows).  Two
representative \gadget\ simulations with $2\times 64^3$ and $2\times
256^3$ particles are shown in the left two columns.  The \enzo\
simulations shown in the right two columns both have a maximum dynamic
range of $\Le = 4096$ and use $128^3$ dark matter particles with a
$128^3$ root grid.  They differ in that the simulation in the
rightmost column uses the \ppm\ hydrodynamic method, while the other column
uses the \zeus\ method.

The gas is initialized at $z=99$ at a temperature of $140\,{\rm K}$
and cools as it adiabatically expands. The gas should follow the
adiabatic relation until it undergoes shock heating, so one expects
that there should be very little entropy production until $z \sim 30$,
because the first gravitationally-bound structures are just beginning
to form at this epoch.  Gas that reaches densities of a few times the
cosmic mean is not expected to be significantly shocked; instead, it
should increase its temperature only by adiabatic compression.  This
is true for \gadget\ and \enzo/\ppm, where almost all of the
gas maintains its initial entropy, or equivalently, it stays on its
initial adiabat. At $z=30$, only a very small amount of high-density
gas departs from its initial entropy, indicating that it has undergone
some shock heating.  However, in the \enzo/\zeus\ simulation, a much
larger fraction of gas has been heated to higher temperatures. In
fact, it looks as if essentially {\em all} overdense gas has increased
its entropy by a non-negligible amount. We believe this is most likely
caused by the artificial viscosity implemented in the \zeus\
method, a point we will discuss further in Section~\ref{viscosity}.

As time progresses, virialized halos and dark matter filaments form,
which are surrounded by strong accretion shocks in the gas and are
filled with weaker flow shocks \citep{Ryu03}. The distribution of gas
then extends towards much higher entropies and densities.  However,
there is still a population of unshocked gas, which can be nicely seen
as a flat constant entropy floor in all the runs until $z=10$.
However, the \enzo/\zeus\ simulation largely loses this feature by
$z=3$, reflecting its poor ability to conserve entropy in unshocked
regions. On the other hand, the \gadget\ `entropy conserving'
SPH-formulation preserves a very well defined entropy floor down to
$z=3$. The result of \enzo/\ppm\ lies between that of \gadget\
and \enzo/\zeus\ in this respect.  The $1 K$ temperature floor
in the \enzo\ code results in an artificial increase in the entropy
``floor'' in significantly underdense gas at $z=3$.  

Perhaps the most significant difference between the simulations lies
however in the `bimodality' that \enzo/\ppm\ develops in the
density-entropy phase space. This is already seen at redshift $z=10$,
but becomes clearer at $z=3$.  While \enzo/\zeus\ and \gadget\ show a
reservoir of gas around the initial entropy with an extended
distribution towards higher density and entropy, \enzo/\ppm\
develops a second peak at higher entropy, i.e.~intermediate density
and entropy values are comparatively rare.  The resulting bimodal
character of the distribution is also reflected in a `dip' at 
$\log T\sim 2.0$ seen in the 1-D differential distribution function in
Figures~\ref{fig.diffdf-1} and~\ref{fig.diffdf-2}.

We note that the high-resolution \gadget\ run with $256^3$ particles
exhibits a broader distribution than the $64^3$ run because of its
much larger dynamic range and better sampling, but it does not show
the bimodality seen in the \enzo/\ppm\ run.  We also find that
increasing the dynamic range $\Le$ with a fixed particle number does
not change the overall shape of the distributions in a qualitative
way, except that the gas extends to a slightly higher overdensity when
$\Le$ is increased.


\subsection{Mean gas temperature and entropy}
\label{mte}

\begin{figure*}
\epsscale{2.0}
\plotone{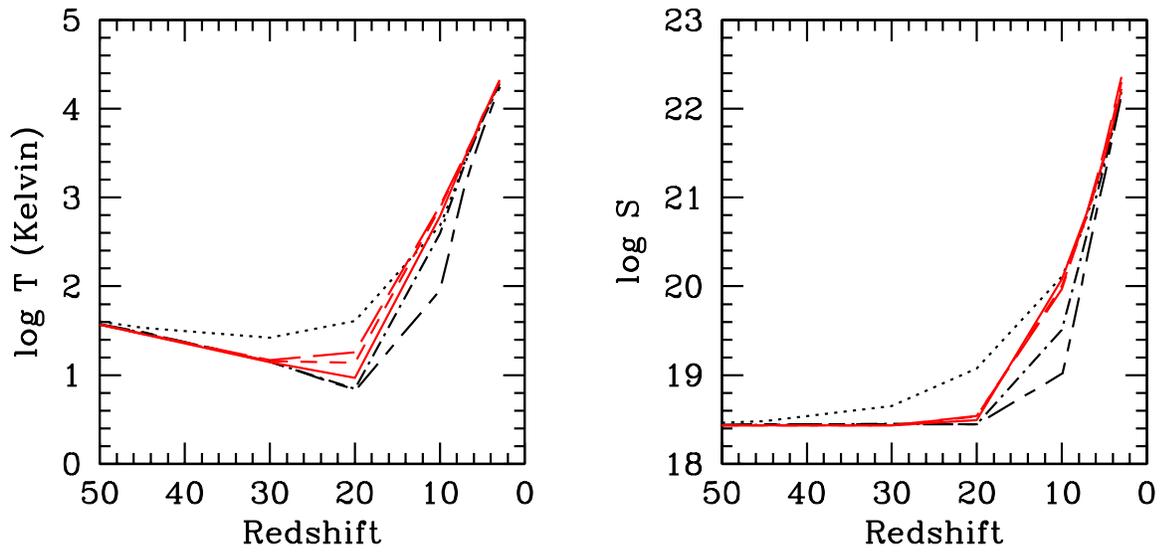}
\caption{Mass-weighted mean gas temperature and entropy for \enzo\ 
and \gadget\ runs as a function of redshift.  
The runs used are the same as those shown in Figures~\ref{fig.diffdf-1}
and~\ref{fig.diffdf-2}.}
\label{fig.gasmeanz}
\end{figure*}

In Figure~\ref{fig.gasmeanz} we show the mass-weighted mean gas
temperature and entropy of the entire simulation box as a function of
redshift.  We compare results for \gadget\ simulations with particle
numbers of $64^3$, $128^3$ and $256^3$, and \enzo\ runs with $64^3$ or
$128^3$ particles for different choices of root grid size and 
hydrodynamic algorithm.

\notetoeditor{Put Fig.12 here.}

In the temperature evolution shown in the left panel of
Figure~\ref{fig.gasmeanz}, we see that the temperature drops until
$z\sim 20$ owing to adiabatic expansion.  This decline in the
temperature is noticeably slower in the \enzo/\zeus\ runs compared
with the other simulations, reflecting the artificial heating seen
in \enzo/\zeus\ at early times. After $z=20$ structure formation and
its associated shock heating overcomes the adiabatic cooling and the
mean temperature of the gas begins to rise quickly.  While at
intermediate redshifts ($z\sim 40-8$) some noticeable differences among
the simulations exist, they tend to converge very well to a common
mean temperature at late times when structure is well developed.  In
general, \enzo/\ppm\ tends to have the lowest temperatures,
with the \gadget\ SPH-results lying between those of \enzo/\zeus\ and
\enzo/\ppm.

In the right panel of Figure~\ref{fig.gasmeanz}, we show the evolution
of the mean mass-weighted entropy, where similar trends as in the mean 
temperature can be observed. We see that a constant initial entropy 
($\log S_{\rm init}=18.44$) is preserved until $z\sim 20$ in 
\enzo/\ppm\ and \gadget.
However, an unphysical early increase in mean entropy is observed in
\enzo/\zeus. The mean entropy quickly rises after $z=20$ owing to
entropy generation as a results of shocks occurring during structure
formation.

Despite differences in the early evolution of the 
mean quantities calculated we find it
encouraging that the global mean quantities of the simulations agree
very well at low redshift, where temperature and entropy converge
within a very narrow range.  At high redshifts the majority of gas
(in terms of total mass) is in regions which are collapsing but still
have not been virialized, and are hence unshocked.  As we show in 
Section~\ref{viscosity}, the formulations of artificial viscosity used
in the \gadget\ code and in the \enzo\ implementation of the \zeus\ 
hydro algorithm play a significant role in increasing the entropy of 
unshocked gas which is undergoing compression (though the magnitude of
the effect is significantly less in \gadget), which explains why the 
simulations using these techniques have systematically higher mean
temperatures/entropies at early times than those using the PPM technique.  
At late times these mean values are dominated by gas which has already 
been virialized in large halos, and the increase in temperature and 
entropy due to virialization overwhelms heating due to numerical effects.
This suggests that most results at low redshift are probably insensitive 
to the differences seen here during the onset of structure formation 
at high redshift.


\subsection{Evolution of Kinetic Energy}
\label{sec.KEevol}

Different numerical codes may have different numerical errors per 
timestep, which can accumulate over time and results in differences 
in halo positions and other quantities of interest. 
It was seen in the Santa Barbara cluster
comparison project that each code calculated the time versus redshift
evolution in a slightly different way, and overall that resulted in 
substructures being in different positions because the codes were at 
different ``times''.  In our comparison of the halo positions in 
Section~\ref{dmhalo} we saw something similar -- the accumulated error 
in the simulations results in our halos being in slightly different 
locations. Since we do not measure the overall integration error in 
our codes (which is actually quite hard to quantify in an accurate way, 
considering the complexity of both codes) we argue that the 
kinetic energy is a reasonable proxy because the kinetic energy is 
essentially a measure of the growth of structure - as the halos grow 
and the potential wells deepen the overall kinetic energy increases.  
If one code has errors that contribute to the timesteps being 
faster/slower than the other code this shows up as slight differences 
in the total kinetic energy. 

\notetoeditor{Put Fig.13 here.}

\begin{figure*}
\begin{center}
\resizebox{9.5cm}{!}{\includegraphics{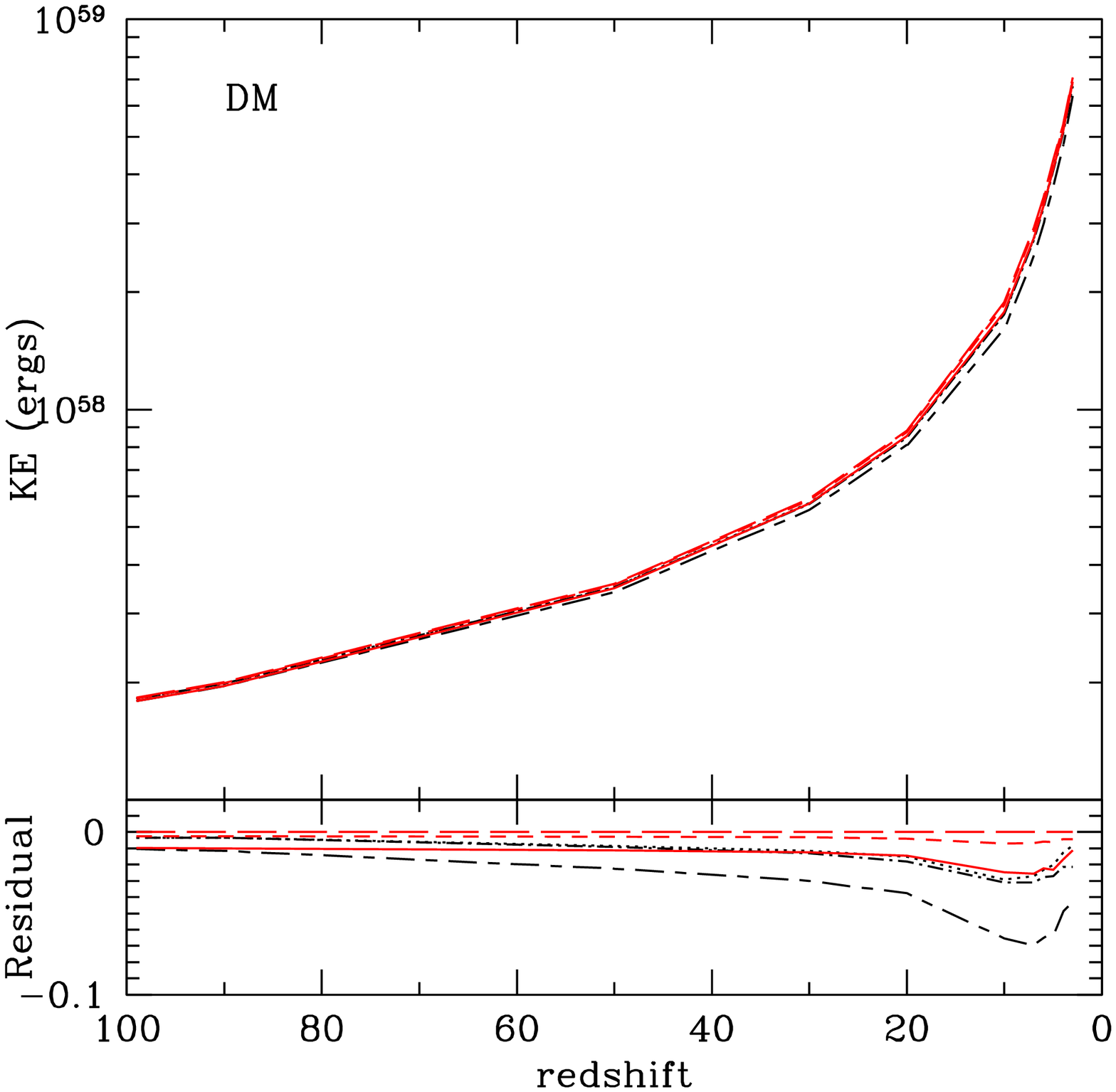}}
\resizebox{9.5cm}{!}{\includegraphics{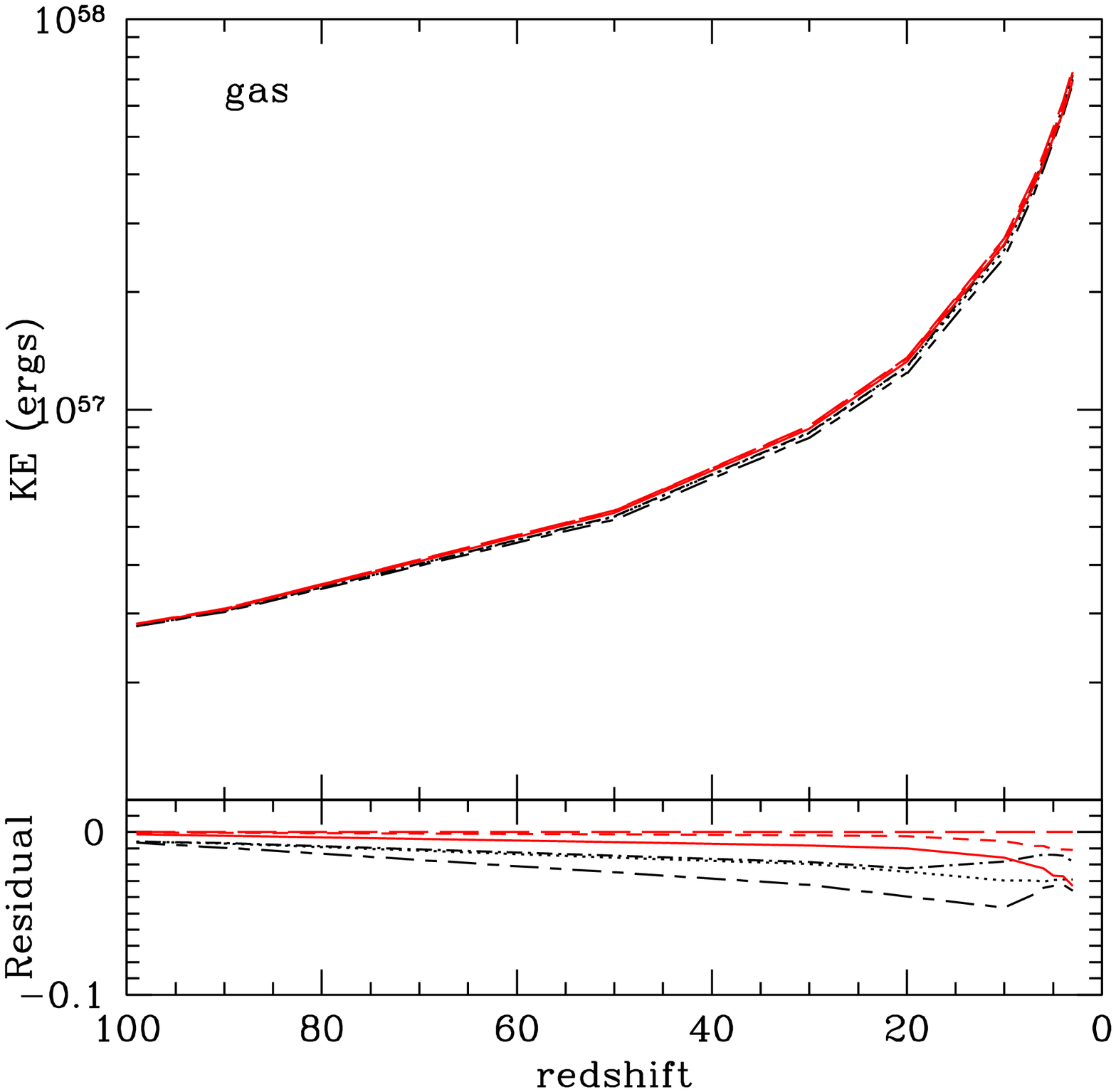}}
\caption{Kinetic energy of dark matter and gas as a function of redshift, 
and the residuals in logarithmic units with respect to the 256$^3$ particle 
Gadget run (red long-dashed line) is shown in the bottom panels.  
Red short-dashed line is for \gadget\ 128$^3$ particle run, and red solid line
is for \gadget\ 64$^3$ particle run. Black lines are for Enzo runs:
128g128dm PPM, lowod (dot-short dash), 128g128dm Zeus (dotted),
64g64dm PPM, lowod (short dash-long dash).
}
\label{fig.KEevol}
\end{center}
\end{figure*}

In Figure~\ref{fig.KEevol} we show the kinetic energy (hereafter KE) of dark
matter and gas in \gadget\ and \enzo\ runs as a function of redshift. As
expected, KE increases with decreasing redshift.  In the bottom panels, the
residuals with respect to the \gadget\ 256$^3$ particle run is shown in
logarithmic units (i.e., $\log$(KE$_{\rm others}$) - $\log$(KE$_{256}$) ).
Initially at $z=99$, \gadget\ and \enzo\ runs agree to within a fraction of a
percent within their own runs with different particle numbers. The
corresponding \gadget\ and \enzo\ runs with the same particle/mesh number
agree within a few percent. These differences may have been caused by the
numerical errors during the conversion of the initial conditions and the
calculation of the KE itself.  It is reasonable that the runs with a larger
particle number result in a larger KE at both early and late times, because
the larger particle number run can sample the power spectrum to a higher
wavenumber, therefore having more small-scale power at early times and more
small-scale structures at late times.  The 64$^3$ runs both agree with each
other at $z=99$, and overall have about 1\% less kinetic energy than the
256$^3$ run. At the same resolution, \enzo\ runs show up to a few percent less
energy at late times than \gadget\ runs, but their temporal evolutions track
each other closely.


\subsection{The gas fraction in halos}
\label{sec.gasfrac}

The content of gas inside the virial radius of dark matter halos is of
fundamental interest for galaxy formation. Given that the Santa
Barbara cluster comparison project hinted that there may be a
systematic difference between Eulerian codes (including AMR)
 and SPH codes (\enzo\ gave slightly 
higher gas mass fraction compared to SPH runs at the virial radius), 
we study this property in our set of simulations.

In order to define the gas content of halos in our simulations we
first identify dark matter halos using a standard friends-of-friends
algorithm. We then determine the halo center to be the center of mass
of the dark matter halo
and compute the ``virial radius'' for each halo using 
Equation~(24) of \citet{Barkana} with the halo mass given by the 
friends-of-friends algorithm. This definition is
independent of the gas distribution, thereby freeing us from
ambiguities that are otherwise introduced owing to the different
representations of the gas in the different codes on a mesh or with
particles. Next, we measure the gas mass within the
virial radius of each halo. For \gadget, we can simply count the SPH
particles within the radius. In \enzo, we include all cells whose centers
are within the virial radius of each halo.  Note that small inaccuracies
can arise because some cells may only partially overlap with the virial 
radius.  However, in significantly overdense regions the cell sizes are
typically much smaller than the virial radius, so this effect should
not be significant for large halos.

\notetoeditor{Put Fig.14 here.}

\begin{figure*}
\begin{center}
\resizebox{9cm}{!}{\includegraphics{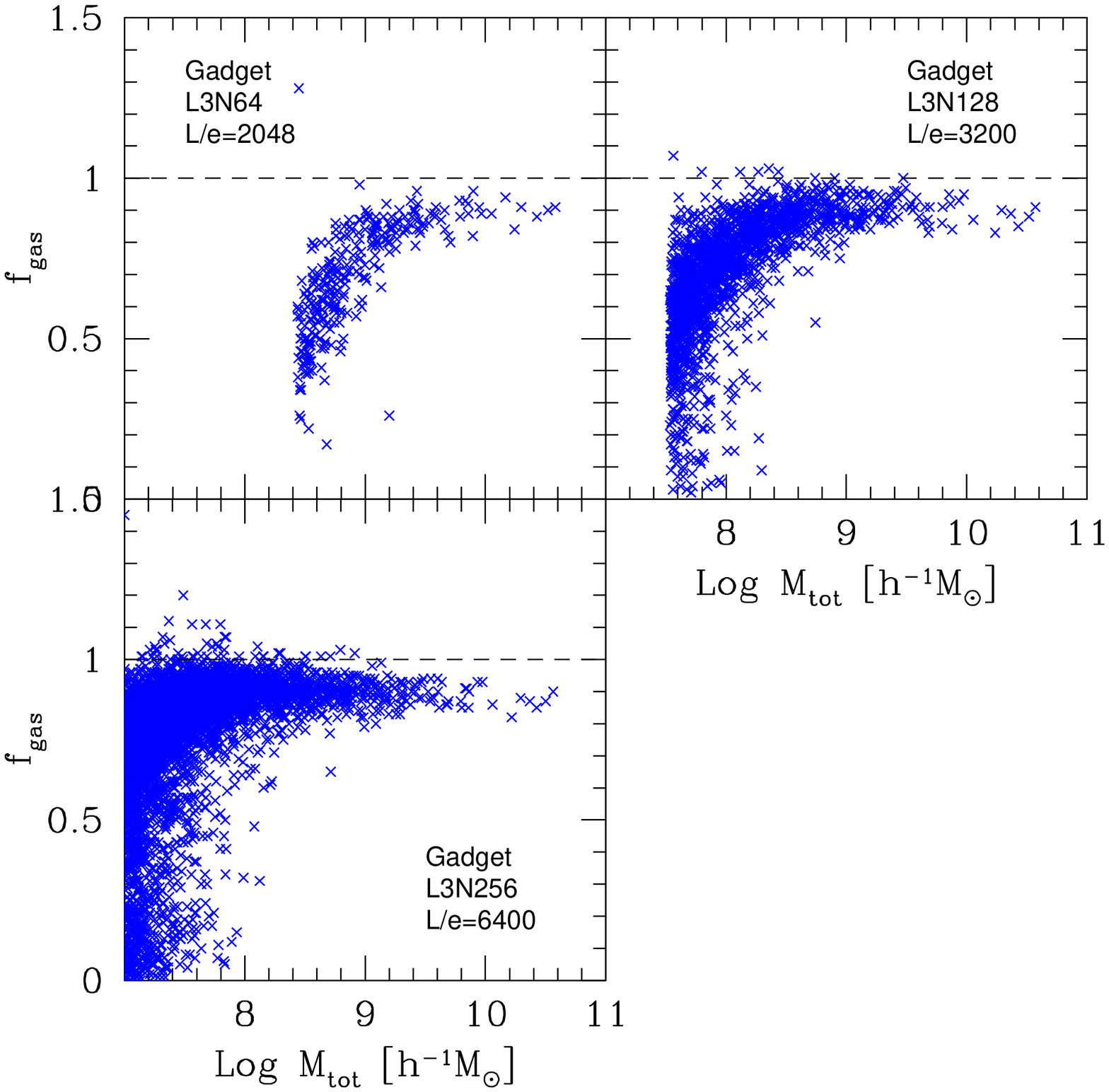}}
\resizebox{10cm}{!}{\includegraphics{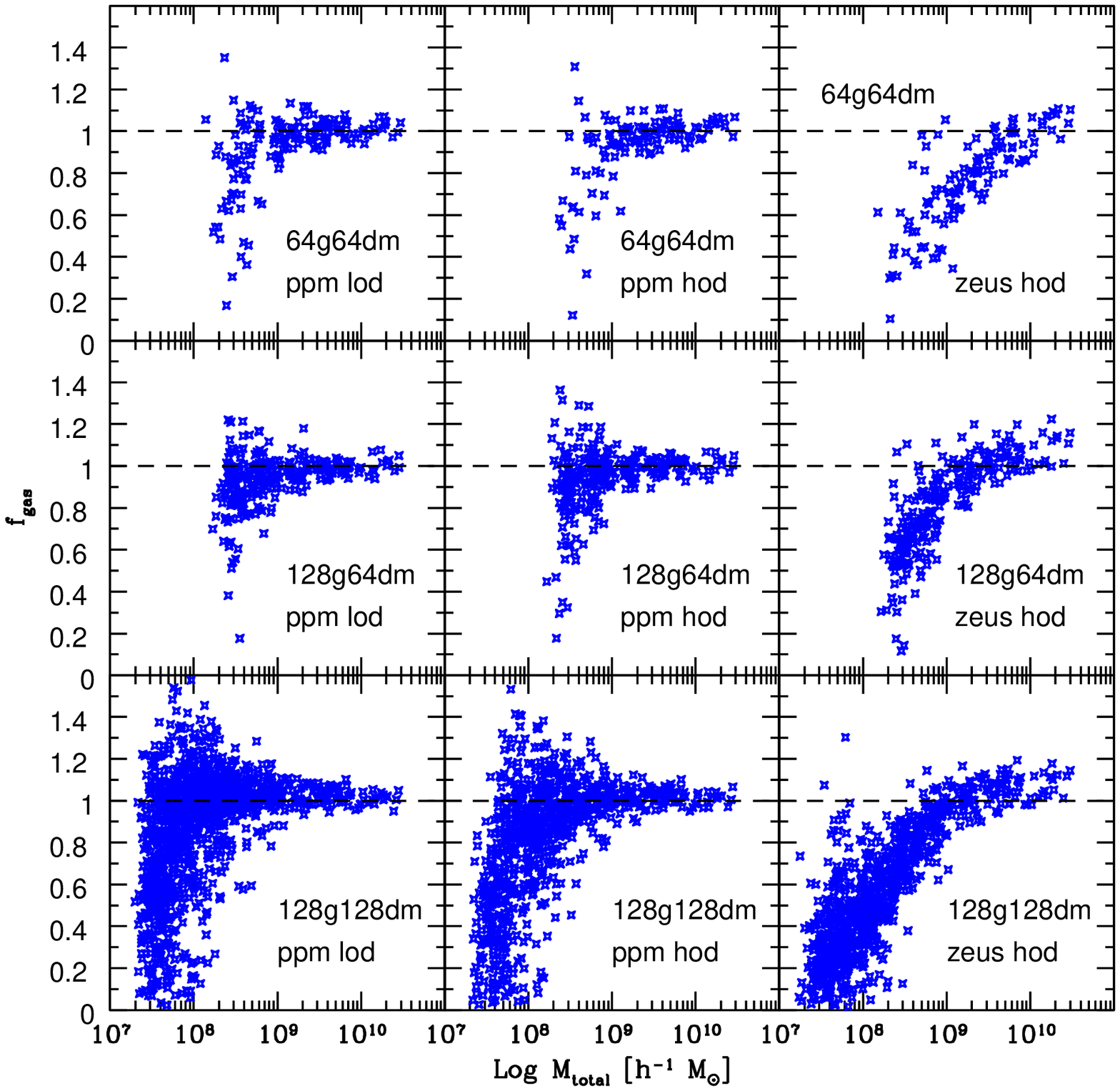}}
\caption{Gas mass fraction normalized by the universal baryon mass fraction 
$f_{\rm gas} = (M_{\rm gas}/M_{\rm tot}) / (\Omega_b/\Omega_m)$ is shown.
The top 3 panels are for \gadget\ runs with $2\times 64^3$,  ($\Le=2048$), 
$2\times 128^3$ ($\Le=3200$), and $2\times 256^3$ ($\Le=6400$) particles.
The bottom panels are for the \enzo\ runs with $64^3$ or $128^3$ grid,
and $64^3$ or $128^3$ dark matter particles. The \zeus\ hydrodynamics
method is used for one set of the \enzo\ simulations (right column) and the
\ppm\ method is used for the rest.
All \enzo\ runs have $\Le=4096$.}
\label{fig.gasmr}
\end{center}
\end{figure*}

In Figure~\ref{fig.gasmr} we show the gas mass fractions obtained in
this manner as a function of total mass of the halos, with the values
normalized by the universal mass fraction $f_{\rm gas} \equiv (M_{\rm
gas}/M_{\rm tot}) / (\Omega_b/\Omega_m)$.  The top three panels show
results obtained with \gadget\ for $2\times 64^4$, $2\times 128^3$,
and $2\times 256^3$ particles, respectively.  The bottom 9 panels
show \enzo\ results with 64$^3$ and 128$^3$ root grids.
Simulations shown in the right column use the \zeus\ hydro algorithm 
and the others use the \ppm\
algorithm.  All \enzo\ runs shown have $64^3$ dark matter particles, except 
for the bottom row which uses $128^3$ particles.  The \enzo\ simulations in
the top row use a $64^3$ root grid and all others use a $128^3$ root grid.  
Grid and particle sizes, overdensity threshold for refinement and hydro 
method are noted in each panel.

For well-resolved massive halos, the gas mass fraction reaches $\sim
90\%$ of the universal baryon fraction in the \gadget\ runs, and $\sim 100\%$
in all of the \enzo\ runs. There is a hint that the \enzo\ runs 
seem to give values a bit higher than the universal fraction, particularly for
runs using the \zeus\ hydro algorithm.  This behavior is
 consistent with the findings of the Santa Barbara 
comparison project.  Given the small size of our sample, it is 
unclear whether this difference is really significant.  However, there is a
clear systematic difference in baryon mass fraction between \enzo\ and \gadget\
simulations.  Examining the mass fraction of simulations to successively 
larger radii show that the \enzo\ simulations are consistently close to 
a baryon mass fraction of unity out to several virial radii, and the gas 
mass fractions for \gadget\ runs approaches unity at radii larger than 
twice the virial radius of a given halo.

The systematic difference between \enzo\ and \gadget\ calculations, even for 
large masses, is also somewhat reflected in the results of \citet{Kravtsov05}.  They perform simulations of galaxy clusters done using adiabatic gas 
and dark matter dynamics with their adaptive mesh code and \gadget.  
At $z=0$ their results for the baryon fraction of gas within the virial 
radius converge to within a few percent between the two codes, with the 
overall gas fraction being slightly less than unity.  It is interesting 
to note that they also observe that the AMR code has a higher overall 
baryon mass fraction than \gadget, though still slightly less 
than what we observe with our \enzo results.

Note that the scatter of the baryon fraction seen for 
halos at the low mass end is a resolution effect. This can 
be seen when comparing the three panels with the \gadget\ results. As
the mass resolution is improved, the down-turn in the baryon fraction
shifts towards lower
mass halos, and the range of halo masses where values near the
universal baryon fraction are reached becomes broader.
The sharp cutoff in the distribution of the points corresponds to 
the mass of a halo with 32 DM particles.

\notetoeditor{Put Fig.15 here.}

\begin{figure*}
\epsscale{1.8}
\plotone{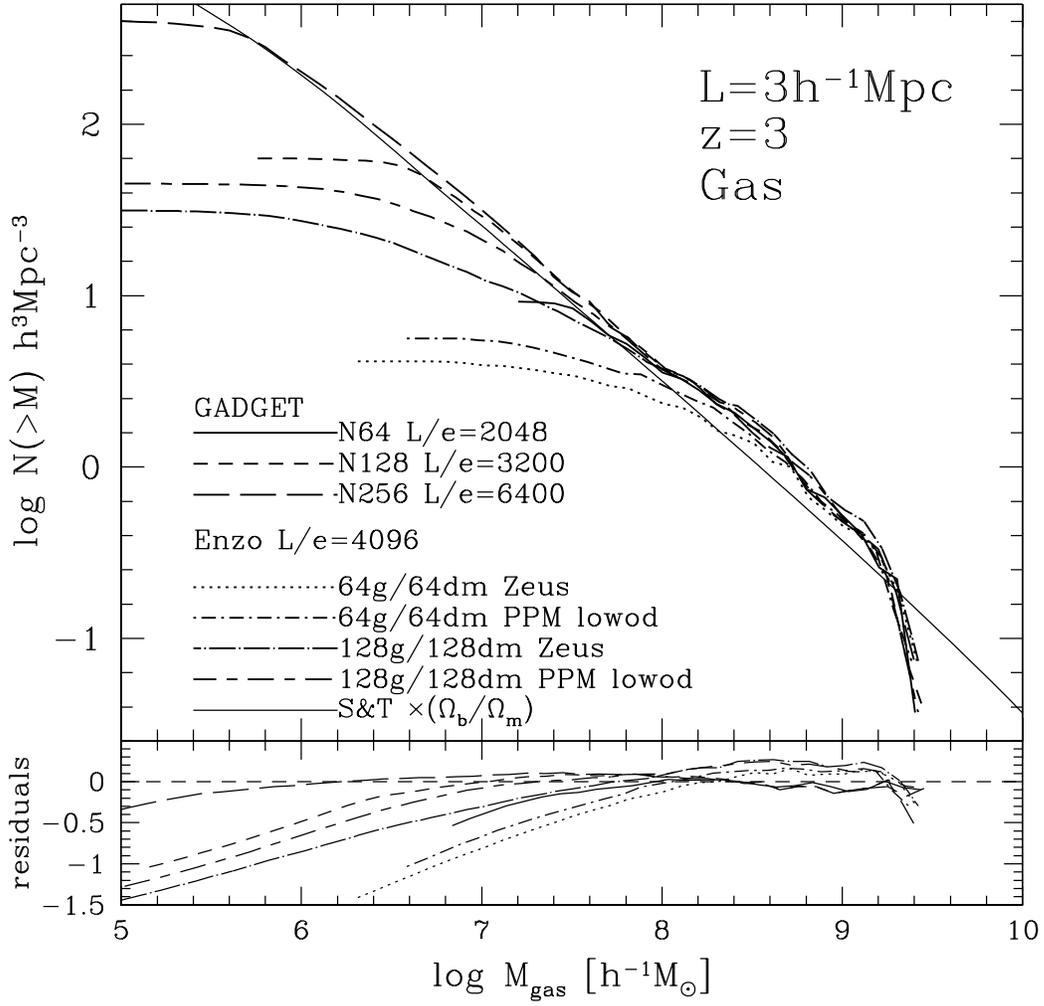}
\caption{Cumulative halo gas mass function at $z=3$. For reference,
the solid black line is the \citet{Sheth99} mass function 
multiplied by the universal baryon mass fraction $\Omega_b/\Omega_m$.
In the bottom panel, the residuals in logarithmic scale with respect 
to the Sheth-Tormen mass function are shown for each run 
(i.e., $\log$(N[$>$M])$ - \log$(S\&T)).
}
\label{fig.gasmf}
\end{figure*}

It is also interesting to compare the cumulative mass function of gas
mass in halos, which we show in Figure~\ref{fig.gasmf} for adiabatic runs. 
This can be viewed as a combination of a measurement of the DM halo mass 
function and the baryon mass fractions. In the lower panel, the residuals
in logarithmic scale are shown for each run with respect to the 
\citet{Sheth99} mass function (i.e., $\log$(N[$>$M])$ - \log$(S\&T)).

As with the dark matter halo mass function, the gas mass functions
agree well at the high-mass end over more than a decade of mass, but
there is a systematic discrepancy between AMR and SPH runs at the
low-mass end of the distribution. While the three SPH runs with
different gravitational softening agree well with the expectation
based on the Sheth \& Tormen mass function and an assumed universal
baryon fraction at $M_{\rm gas} < 10^8\himsun$, the \enzo\ run with 
$64^3$ root grid and $64^3$ DM particles has fewer halos.
Similarly, the \enzo\ run with $128^3$ grid and $128^3$ DM particles
has fewer low mass halos at $M_{\rm gas} < 10^7\himsun$ compared
to the \gadget\ $128^3$ DM particle run. Convergence with the SPH 
results for \enzo\ requires the use of a root grid with spatial 
resolution twice that of the initial mean interparticle separation, 
as well as a low-overdensity refinement criterion. 
We also see that the PPM method results in a  better gas mass function 
than the ZEUS hydro method at the low-mass end for the same number
of particles and root grid size.


\section{The role of artificial viscosity}
\label{viscosity}

\notetoeditor{Put Fig.16 here.}

In Section \ref{2d-dist-S} we found that slightly overdense gas in
\enzo/\zeus\ simulations shows an early departure from the adiabatic
relation towards higher temperature, suggesting an unphysical entropy
injection. In this section we investigate to what extent this effect can be
understood as a result of the numerical viscosity built into the
\zeus\ hydrodynamic algorithm.  As the gas in the pre-shocked universe
begins to fall into potential wells, this artificial viscosity causes
the gas to be heated up in proportion to its compression, potentially
causing a significant departure from the adiabat even when the shock
has not occurred yet; i.e.~when the compression is only adiabatic.

\begin{figure*}
\epsscale{1.7}
\plotone{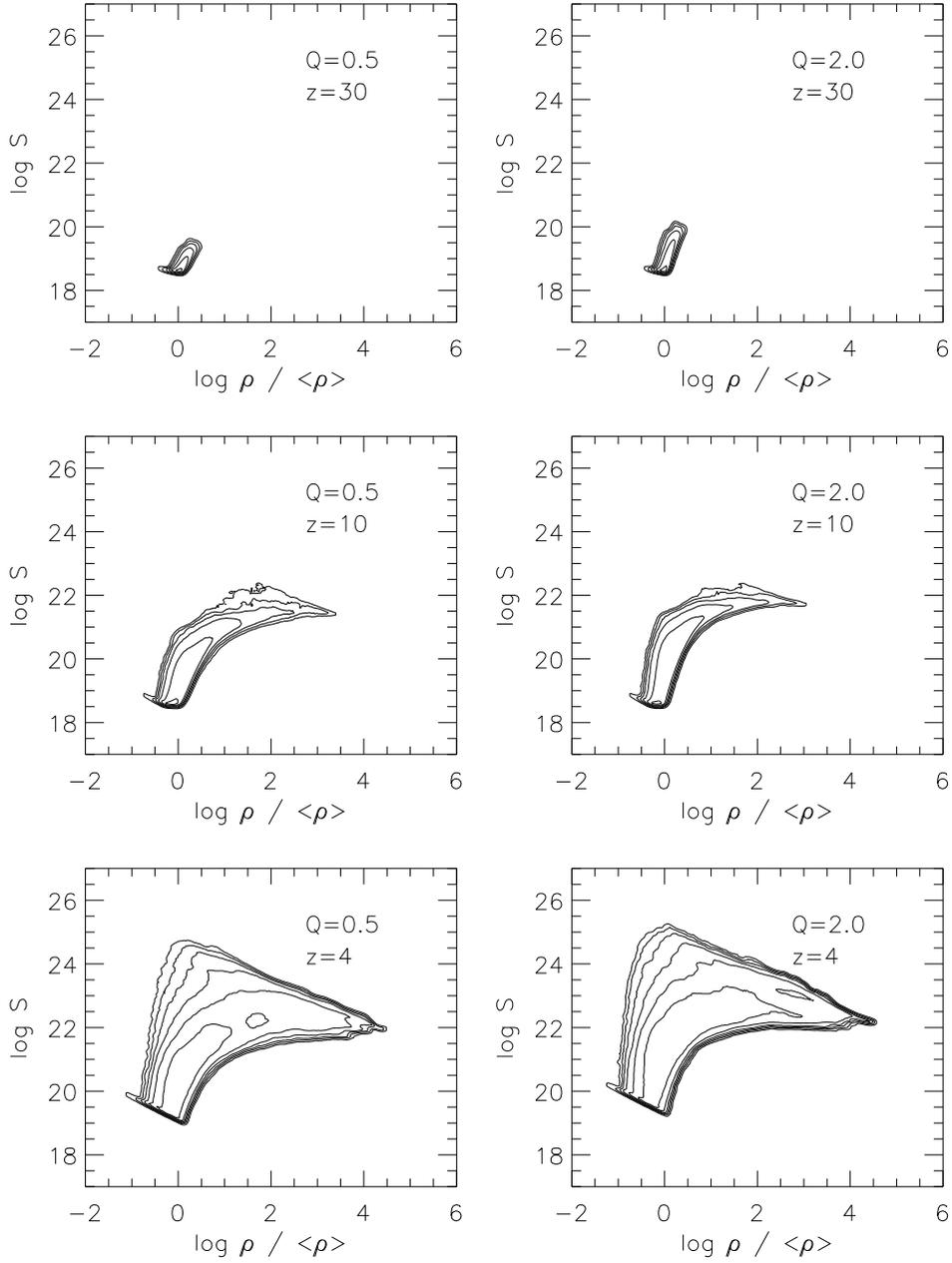}
\caption{Two-dimensional distribution functions of gas entropy vs. gas
overdensity for two \enzo\ runs performed with the \zeus\
hydrodynamics algorithm, varying with redshift.  Rows
correspond to (top to bottom) $z=30$, 10 and 3.  In each panel, six
contours are evenly spaced from 0 to the maximum value in equal logarithmic
scale.  Two different values of the \zeus\ artificial viscosity parameter 
are used: $Q_{\rm AV} = 0.5$ (left column) and $Q_{\rm AV} = 2.0$ 
(right column).  Both runs use $64^3$ dark matter particles and a $64^3$ 
root grid and have a maximum spatial resolution of $\Le = 4096$. 
The standard value of the artificial viscosity parameter is 
$Q_{\rm AV} = 2.0$.}
\label{fig.2d_S_zeusav}
\end{figure*}

This effect is demonstrated in Figure~\ref{fig.2d_S_zeusav}, where we
compare two-dimensional entropy--overdensity phase space diagrams for
two \enzo/\zeus\ where the strength of the artificial viscosity was
reduced from its ``standard'' value of $Q_{\rm AV}=2.0$ to $Q_{\rm
AV}= 0.5$. These runs used $64^3$ dark matter particles and $64^3$
root grid, and the $Q_{\rm AV}=2.0$ corresponds to the case shown
earlier in Figure~\ref{fig.2d_S_evol}.

Comparison of the \enzo/\zeus\ runs with $Q_{\rm AV}=0.5$ and 2.0
shows that decreasing $Q_{\rm AV}$ results in a systematic decrease of
the unphysical gas heating at high redshifts.  Also, at $z=4$ the
$Q_{\rm AV}=0.5$ result shows a secondary peak at higher density, so
that the distribution becomes somewhat more similar to the \ppm\ result.
Unfortunately, a strong reduction of the artificial viscosity in the
\zeus\ algorithm is numerically dangerous because the discontinuities
that can appear owing to the finite-difference method are then no longer
smoothed sufficiently by the artificial viscosity algorithm, which can
produce unstable or incorrect results.

An artificial viscosity is needed to capture shocks when they occur
in both the \enzo/\zeus\ and \gadget\ SPH scheme. This in itself
is not really problematic, provided the artificial viscosity is very
small or equal to zero in regions without shocks. In this respect,
\gadget's artificial viscosity behaves differently from that of
\enzo/\zeus.  It takes the form of a pairwise repulsive force that is
non-zero only when Lagrangian fluid elements approach each other in
physical space. In addition, the strength of the force depends in a
non-linear fashion on the {\em rate of compression} of the
fluid. While even an adiabatic compression produces some small amount
of (artificial) entropy, only a compression that proceeds {\em
rapidly} with respect to the sound-speed, as in a shock, produces
entropy in large amounts.  This can be seen explicitly when we analyze
equations (\ref{sphviscosity}) and (\ref{entropyproduction}) for the
case of a homogeneous gas which is uniformly compressed. For
definiteness, let us consider a situation where all separations shrink
at a rate $q = \dot r_{ij}/r_{ij} < 0$, with $\vec {\nabla} \cdot
\vec{v} = 3\,q$. It is then easy to show that the artificial viscosity
in \gadget\ produces entropy at a rate
\begin{equation}
\frac{{\rm d} \log A_i}{{\rm d} \log \rho_i} = 
\frac{\gamma-1}{2} \alpha
\left[\frac{-q\, h_i} {c_i}  
+ 2 \left( \frac{q \, h_i} {c_i}\right)^2\right].
\end{equation}
Note that since we assumed a uniform gas, we here have $h_i = h_{ij}$,
$c_i= c_{ij}$, and $\rho_i = \rho_{ij}$.  We see that only if the
compression is fast compared to the sound-crossing time across the
typical spacing of SPH particles, i.e.~for $|q| > c_i/h_i$, a
significant amount of entropy is produced, while slow (and hence
adiabatic) compressions proceed essentially in an isentropic
fashion. On the other hand, the artificial viscosity implemented in
\enzo/\zeus\ produces entropy irrespective of the sound-speed,
depending only on the compression factor of the gas.

\notetoeditor{Put Fig.17 here.}

We have also investigated the pre-shock entropy generation in
\enzo/\zeus\ using another simple test, the collapse of a
one-dimensional Zel'dovich pancake.  The initial conditions of this
test are simple and described in full detail in \citet{Bryan95}.  A
one-dimensional simulation volume
 is set up in an expanding coordinate system in a flat
cosmology with an initially sinusoidal density 
perturbation with a peak at $x=0.5$ and a corresponding perturbation
in the velocity field with nodes at $x=0.0$, 0.5, and 1.0.  
A temperature perturbation is added such that gas entropy is
constant throughout the volume.

\begin{figure*}
\epsscale{2.0}
\plotone{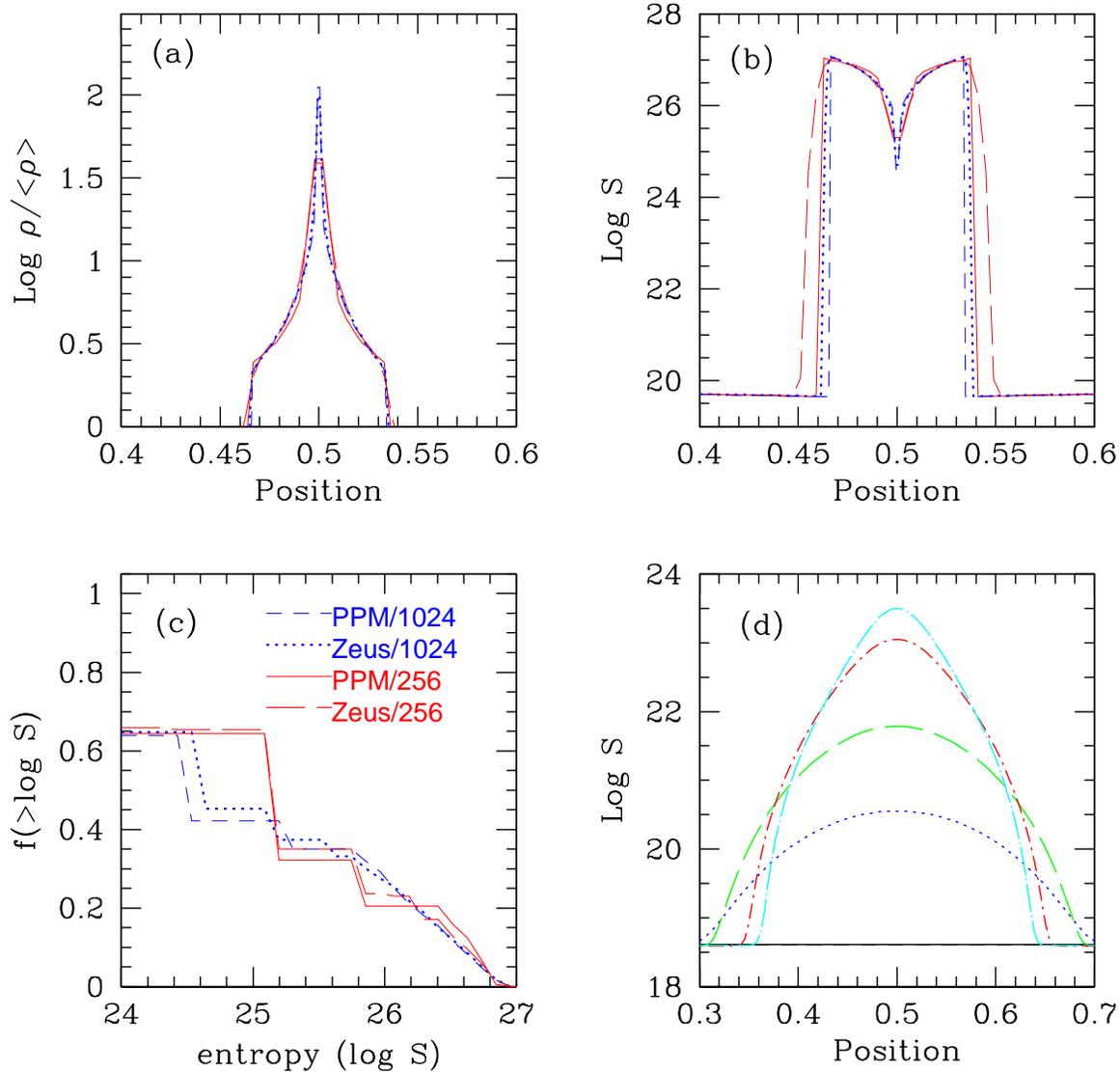}
\caption{Final results of the Zel'dovich pancake test. 
{\it Panel (a)}: Log baryon overdensity vs. position.  
{\it Panel (b)}: Log entropy vs. position.  
{\it Panel (c)}: Cumulative mass-weighted entropy distribution function.  
{\it Panel (d)}: Pre-shock entropy evolution in the 256-cell Zeus run. 
All Zel'dovich pancake simulations are performed in one dimension  
using the \enzo\ code. 
{\it Panels (a), (b), (c)}:
The line types are for \ppm/1024 (short-dashed), \ppm/256 (solid), 
\zeus/1024 (dotted), and \zeus/256 (long dashed) where 256 and 1024
are the number of cells used in the calculation.  All data is at the
end of the run $(z=0)$.  
{\it Panel (d)}:  Entropy evolution of the 256-cell \zeus\
and \ppm\ runs for redshifts $z=20$ (black), 10 (blue dotted), 
5 (green long-dashed), 2.5 (red dot-dashed) and 2 (cyan dot-long-dashed).
All \ppm\ results overlay the \zeus\ initial conditions $(z=20)$.
Note that the x-axis range for panel (d) is different from that of
panels (a) and (b).
}
\label{fig.zp_final}
\end{figure*}

In Figure~\ref{fig.zp_final} we show the density and
entropy profiles as a function of
position, at a time when the non-linear collapse of the pancake is
well underway.  We also show the pre-shock evolution of the
entropy profile for both algorithms.
We compare runs using 256 and 1024 grid cells with
both the \zeus\ and {\small PPM} formulations of \enzo.  

As the matter falls in onto the density peak at $x=0.5$, accretion
shocks on either side form, clearly marked by the jumps in density,
entropy, and temperature.  Note that the dip in the temperature at
$x=0.5$ is physical -- the gas sitting there is unshocked and only
adiabatically compressed, and therefore has relatively low temperature.
Reassuringly, both the \zeus\ and \ppm\ hydrodynamical methods
reproduce the qualitative behavior of the Zel'dovich pancake quite
well, but there are also some systematic differences at a given
resolution. This can be seen most clearly
 in the mass-weighted cumulative entropy
distribution in the bottom left panel of Figure~\ref{fig.zp_final}.
We see that the \enzo/\zeus\ calculations show
 a broader distribution than \enzo/\ppm\ for a given spatial resolution.
This can be interpreted as another sign of pre-shock entropy
generation by the artificial viscosity in \zeus. In contrast, the
Riemann solver used in \ppm\ can capture shocks such that they
are resolved as true discontinuities, which avoids this problem.  

More concrete evidence of spurious entropy generation in the
artificial viscosity-based scheme can be seen by examining the pre-shock
evolution of entropy in these simulations (as seen in panel (d) of 
Figure~\ref{fig.zp_final}).  No entropy should be generated before
the twin shocks form to the left and right of $x=0.5$ (as can be
seen in panel (b) of the same figure).  The simulations using \ppm\
(black solid line in panel d) produce no spurious entropy.  The simulations
using the \zeus\ scheme, however, produce significant amounts of entropy
in the infalling (but unshocked) gas.  Note that the magnitude of the
entropy generation is relatively small compared to the final entropy
produced in the shocks (as seen in panel ($b$)), but the values are 
still significant.

While this test showed only comparatively small differences between the
different methods, it is plausible that the effects of pre-shock
entropy generation become much more important in three-dimensional
cosmological simulations, where galaxies form hierarchically through
complicated merger processes that involve extremely complex shock
patterns. We thus speculate that this effect may be the key reason for
the systematic differences between the \enzo/\ppm\ runs and the \zeus\
and \gadget\ simulations.


\section{Timing \& memory usage}
\label{timing}

An important practical consideration when assessing the relative
performance of computational methods or simulation codes is the
amount of computational resources they require to solve a given
problem. Of primary importance are the total amount of
memory and the CPU time that is needed.  However, it is not always
easy to arrive at a meaningful comparison, particularly for very
different methods such as AMR and SPH.  For example, the variable
number of grid cells owing to the adaptive nature of AMR is an 
important complication, making the number of resolution elements 
change over time, while the particle number stays constant in the 
SPH method.  An additional layer of complexity is added when 
considering parallel codes.  The parallelization strategies that 
are used for AMR applications can be significantly different 
than those used in SPH codes, and the performance of an individual 
simulation code can heavily depend on the specific computer architecture 
and implementation of MPI (or other software used for parallelization) 
chosen. Therefore we caution the readers to take all of the timing 
information discussed in this section as results for a particular 
problem setup and machine architecture, and not to extrapolate 
directly to different types of cosmological simulations 
(e.g., with cooling and star formation) and machines.

\subsection{Initial comparison on a distributed memory machine}

When we started this project, we initially performed our comparison
runs on the IA-64 Linux cluster Titan at the National Center for 
Supercomputing Applications (NCSA). It had 134 dual processor nodes
with 800 MHz Intel Itanium 1 chips, 2.5 GB memory per node, and 
Myrinet 2000 network interconnect. Our initial comparison on Titan
showed that the \gadget\ code was faster than \enzo\ by a factor of 
40 (15) for a 64$^3$ (128$^3$) particle DM-only run when \enzo\ was 
using a low overdensity criteria for grid refinement. 
The low overdensity refinement criterion 
was required for \enzo\ in order to obtain a DM halo mass function 
comparable to that of \gadget\ at low-mass end. \gadget\ used  
a factor of 18 (4) less amount of memory than \enzo\ for a 64$^3$ 
(128$^3$) particle DM-only run. For the adiabatic runs, \gadget\
was faster than \enzo\ by a factor of 2.5 for a $64^3$ DM particles 
and 64$^3$ gas particles (a 64$^3$ root grid for \enzo). 
A \gadget\ run with 128$^3$ dark matter and gas particles completed 
8 times faster than an \enzo\ simulation with a 128$^3$ root grid and 
64$^3$ DM particles.  
These performance results were gathered using Linux-based 
Beowulf-style clusters with relatively slow inter-node
communication networks. Since the AMR code performs load balancing
by passing grids between processors, it was expected that the performance
of \enzo\ would improve on a large shared-memory machine. 
The disparity is most significant for DM-only simulations, so 
improvement of the \enzo\ N-body solver could significantly increase 
the performance of the AMR code.

\subsection{More recent comparison on a shared memory machine}

During the course of this comparison study, both \gadget\ and \enzo\
evolved, and the performance of both codes have greatly improved. 
Therefore, we repeated the performance comparison with our updated codes 
using the IBM DataStar machine at the San Diego Supercomputing 
Center.\footnote{$http://www.sdsc.edu/user\_services/datastar/$}  
The portion of the machine used for these timing tests is composed of 
176 IBM p655 compute nodes, 
each of which has eight 1.5 GHz IBM Power4 processors.  These processors are
super-scalar, pipelined 64 bit chips which can execute up to 8 
instructions per clock cycle and up to four floating point operations
per clock cycle, with a theoretical peak performance of 6.0 GFlop per 
chip.  Processors in a single node share a total of 16 GB of memory.
All nodes are connected by an IBM Federation switch, which provides
processor-to-processor bandwidth of approximately 1.4 GB/s with 
8 microsecond latency when using IBM's MPI library.  Each node is
directly connected to a parallel filesystem through a Fibre Channel
link.  

We first compare the series of dark matter-only runs discussed in
Section~\ref{dm}.  A \gadget\ simulation with $64^3$ dark matter particles 
takes total wall-clock time of 225 seconds on 8 cpus (total 1800 seconds 
CPU time) and requires 270 MB of memory.  24\% of the total 
computational time was spent doing interprocessor message-passing.  
The corresponding \enzo\ simulation 
with $64^3$ particles and a $64^3$ root grid requires 1053 seconds 
on 8 cpus (total 8424 seconds CPU time) when refining on a dark matter 
overdensity of 2.0, and requires 1.21 GB of memory total. 
34\% of the total computational time was spent in interprocessor 
communication.  This is a factor of 4.7 slower than the corresponding 
\gadget\ simulation, and requires roughly 4.5 times more memory.  
Raising the refinement criteria to a dark matter overdensity of 4.0 
(at a cost of losing low-mass DM halos) reduces the wall clock time to 
261 seconds on 8 processors (total 2088 seconds CPU time) and decreases 
the total amount of memory 
needed to 540 MB, which is comparable to the \gadget\ simulation.  
A $128^3$ DM particle \gadget\ adiabatic run takes a total of 2871 seconds 
to run on 8 cpus (total 22,968 seconds CPU time) and requires 1.73 GB of memory.  
An \enzo\ simulation with $128^3$ particles and a $128^3$ root grid that 
refines on a dark matter overdensity of 2.0 needs approximately 34,028
seconds on 8 processors (total 272,224 CPU seconds) and 5.6 GB of memory. 
This is a factor of 12 slower and 3.2 times more memory than the equivalent
\gadget\ run.  The same calculation run with refinement overdensities 
of 4.0 or 8.0 completes in 13,960 and 3839 seconds, respectively, 
which are factors of 4.9 and 1.3 slower than the equivalent 
\gadget\ run.
The reason for the huge change in computational speeds is due to the low
overdensity threshold used in the first simulation, which results in a 
huge number of grids to be instantiated and a great deal of time to be 
spent regridding the simulation.
Raising the overdensity criteria suppresses the formation of halos at the low
mass end of the mass function, though higher-mass halos are unaffected.
This timing comparison suggests that if one is interested in simulating 
the full spectrum of dark matter halos at a reasonable computational
cost, \gadget\ would be a wiser choice than \enzo\ for this application.
If one was interested in only the high-mass end of the mass function,
the codes have comparable performance.

Comparison of the adiabatic gas + N-body cosmological simulations in 
Section~\ref{baryon} is also quite informative.  The $64^3$ dark matter 
particle/$64^3$ gas particle \gadget\ calculation takes 1839 seconds 
to run on 8 processors (total 14,712 seconds CPU time) and requires 
511 MB of memory.  The equivalent \enzo\ simulation with $64^3$ particles 
and a $64^3$ root grid using the low overdensity refinement
criteria (refining on a baryon overdensity of 4.0 and a dark matter 
overdensity of 2.0) requires 6895 seconds on 8 processors (55,160 seconds 
total) and 2.5 GB of memory.  This is 3.7 times slower and 4.9 times more 
memory than the corresponding \gadget\ run. Raising the overdensity 
thresholds by a factor of two decreases the computational time to 
2168 seconds on 8 processors and the memory required to 1.28 GB.  
The \gadget\ calculation with $128^3$ dark matter
and baryon particles requires 35,879 seconds on 8 cpus (287032 seconds 
total CPU time) and 5.4 GB of memory, and an \enzo\ calculation with 
$128^3$ particles on a $128^3$ root grid which refines on a baryon 
overdensity of 8.0 and a dark matter overdensity of 4.0 requires 
64,812 seconds and 8 GB of memory.  \enzo\ simulations using the PPM 
and Zeus hydro methods require comparable amounts of simulation time.


\section{Conclusions}
\label{discussion}

This paper presents initial results of a comparison of two
state-of-the-art cosmological hydrodynamic codes: \enzo, an Eulerian
adaptive mesh refinement code, and \gadget, a Lagrangian smoothed particle
hydrodynamics code. These codes differ substantially in the way they
compute gravitational forces and even more radically in the way they
treat gas dynamics. In cosmological applications structure formation
is driven primarily by gravity, so a comparison of the hydrodynamical
methods necessarily involves an implicit comparison of the
gravitational solvers as well. In order to at least partially
disentangle these two aspects we have performed both a series of dark 
matter-only simulations and a set of simulations that followed both 
a dark matter and an adiabatic gaseous component.

Our comparison of the dark matter results showed good agreement in
general provided we chose a root grid resolution in \enzo\ at least
twice that of the mean interparticle separation of dark matter
particles together with a relatively conservative AMR refinement
criterion of dark matter overdensity of 2. If less stringent settings
are adopted, the AMR code shows a significant deficit of low mass
halos. This behavior can be readily understood as a consequence of
the hierarchical particle-mesh algorithm used by \enzo\ for computing
gravitational forces, which softens forces on the scale of the mesh
size. Sufficiently small mesh cells are hence required to compete with
the high force-resolution tree-algorithm of \gadget.  In general, we 
find excellent agreement with the results of \citet{Heitmann05}, 
particularly with regards to systematic differences in the power 
spectrum and low-mass end of the halo mass function between mesh 
and tree codes.  Our results are complementary in several ways --
Heitmann et~al. use simulations run with the ``standard'' parameters
for many codes (using the same initial conditions) and then compare 
results without any attempt to improve the quality of agreement, 
whereas we examine only two codes, but systematically vary 
parameters in order to understand how the codes can be made to agree
to very high precision.

Examination of the dark matter substructure in the two most massive
halos in our $128^3$ particle dark matter-only calculations shows that 
while both codes appear to resolve substructure (and obtain substructure 
mass functions that are comparable) there are some differences in the 
number and the spatial distribution of subhalos between the two codes. 
While the origin of these differences are not fully clear, it may be 
due to a lack of spatial (i.e. force) or dark matter mass resolution, 
or possible due in part to systematics in the grouping algorithm used 
to detect substructure.  The observed differences in substructure
are not surprising when one considers how dissimilar the algorithms
that \enzo\ and \gadget\ use to calculate gravitational accelerations
on small scales are, and a further study with much higher resolution 
is necessary. 

We also found broad agreement in most gas quantities we examined in 
simulations which include adiabatic gas evolution, but there were also 
some interesting discrepancies between the different codes and 
different hydrodynamical methods. While the distributions of temperature, 
density, and entropy of the gas evolved qualitatively similarly over time, 
and reassuringly converged to the same mean temperature and entropy 
values at late times, there were clearly some noticeable differences 
in the early evolution of the gas and in the properties of intermediate 
density gas.

In particular, in the \enzo/\zeus\ simulations we found an early
heating of collapsing or compressed gas, caused by injection of
entropy by the artificial viscosity in this code.  This resulted in
substantial pre-shock entropy generation in the \enzo/\zeus\ runs.
While \gadget\ also uses an artificial viscosity to capture shocks,
effects of pre-shock entropy generation are substantially weaker in
this code.  This reflects its different parameterization of artificial 
viscosity, which better targets the entropy production to shocked regions.

Considering the entropy-density distribution in more detail, we found
that \enzo/{\small PPM} calculations show a marked trend towards 
a segregation of gas
into a low-entropy reservoir of unshocked low density gas and a pool
of gas that has been shocked and accumulated high entropy when it
reached higher density regions. Such a bimodality is not apparent in the
\enzo/\zeus\ and \gadget\ runs at $z=3$. Instead, there is a smoother
transition from low- to high-entropy material; i.e.~more gas of
intermediate entropy exists. It is possible that this
intermediate-entropy gas is produced by the artificial viscosity in
pre-shock regions, where entropy generation should not yet take place.
Some supporting evidence for this interpretation is provided by the
fact that the distributions of temperature and entropy of \enzo/\zeus\
become somewhat more similar to those of \enzo/{\small PPM}\ when we
reduce the strength of the artificial viscosity.

Perhaps the most interesting difference we found between the two methods
lies in the baryon fraction inside the virial radius of the halos at $z=3$.  
For well-resolved halos \enzo\ results asymptote to slightly higher than
$100\%$ of the cosmic baryon fraction, independent of the resolution 
and hydro method used (though note that the results using the \zeus\
method appear to converge to a marginally higher value than the
PPM results). This also shows up as an overestimate of gas
mass function $M_{gas}>10^8 \himsun$ compared to the Sheth \& Tormen 
function multiplied by $(\Omega_b/\Omega_M$).  In contrast, 
\gadget\ halos at all resolutions only reach $\sim 90\%$ of the 
cosmic baryon fraction.  This result is not easily understood in terms 
of effects due to artificial viscosity since the \zeus\ method used 
in \enzo\ produces more artificial viscosity than either of the 
other methods, yet the results for the two hydro methods in \enzo\ 
agree quite well.  The systematic difference between \enzo\ and
\gadget\ results in this regime provides an interesting comparison to
\citet{Kravtsov05}, who examine the enclosed gas mass fraction at $z=0$ 
as a function of radius of eight galaxy clusters in adiabatic gas 
simulations done with the ART and \gadget\ codes.  They see that at small 
radii there are significant differences in enclosed gas mass fraction, 
but at distances comparable to the virial radius of the cluster the mass
fractions converge to within a few percent and are overall approximately 
$95\%$ of the universal mass fraction.  It is interesting to note that the 
enclosed gas mass fraction at the virial radius produced by the ART code 
is higher than that of \gadget\ by a few percent, and the ART gas mass 
fraction result would be bracketed by the \enzo\ and \gadget\ results, 
overall.  This suggests that it is not clear that a universal baryon 
fraction of $\sim 100\%$ is predicted by AMR codes, though there seems 
to be a clear trend of AMR codes having higher overall baryon mass fractions 
in halos than SPH codes to, which agrees with the results of \citet{Frenk99}

It is unclear why our results with the \gadget\ code differ from those 
seen in Kravtsov et al. (with the net gas fraction in our calculations 
being approximately $5\%$ lower at the virial radius), though it may be 
due entirely to the difference in regime -- we are examining galaxy-sized 
halos with masses of $\sim 10^9-10^{10} M_{\odot}$ at $z=3$, whereas they 
model $\sim 10^{13} - 10^{14} M_{\odot}$ galaxy clusters at $z=0$. 
Regardless, the observed differences between the codes are significant 
and will be examined in more detail in future work.

It should be noted that the hydrodynamic results obtained for the \gadget\ 
SPH code are typically found to be bracketed by the two different 
hydrodynamic formulations implemented in the AMR code.
This suggests that there is no principle systematic
difference between the techniques which would cause widely differing
results. Instead, the systematic uncertainties within each technique,
for example with respect to the choice of shock-capturing algorithm,
appear to be larger than the intrinsic differences between SPH and AMR
for the quantities of interest in this paper.  
We also note that some of the differences we find in bulk simulation
properties are likely to be of little relevance for actual simulations
of galaxy formation. For example, in simulations including more
realistic physics, specifically a UV background, the low temperature gas 
that is affected most strongly by artificial early
heating in \enzo/\zeus\ will be photoionized and thus heated
uniformly to approximately 10$^4\,{\rm K}$, so that many of the 
differences in temperature and entropy at low overdensity owing to the
choice of hydrodynamical method will disappear. We will investigate
such effects of additional physics in the future.

We have also examined the relative computational performance of the
codes studied here, using metrics such as the total CPU time and
memory consumption. If one simply compares simulations which have the
same number of particles and grid cells at the start of the
simulation, \gadget\ performs better; i.e.~it finishes faster,
uses less memory, and is more accurate at the low-mass end of the halo
mass function. However, much of this difference is caused by the
slowly increasing number of cells used by the AMR code to represent
the gas, while the Lagrangian code keeps the number of SPH particles
constant. If the consumed resources are normalized to the number of
resolution elements used to represent the gas (cells or particles),
they are roughly comparable.  Unfortunately, the lower gravitational
force-resolution of the hierarchical particle-mesh algorithm of \enzo\
will usually require the use of twice as many root grid cells
as particles per dimension for high-accuracy results at the low-mass
end of the mass function, which then
induces an additional boost of the number of needed cells by nearly an
order of magnitude with a corresponding impact on the required
computational resources. As a consequence of this, the gas will be
represented more accurately, and this is hence not necessarily a
wasted effort. However given that the dark matter mass resolution is not
also improved at the same time (unless the DM particle number is also 
increased), it is probably of little help to make progress in the 
galaxy formation problem, where the self-gravity of dark matter is 
of fundamental importance.  It is also true that the relative 
performance of the codes is dependent upon the memory architecture 
and interprocessor communication network of the computer used to 
perform the comparison as we discussed in Section~\ref{timing}.

It is encouraging that, with enough computational effort, it is 
possible to achieve the same results using both the \enzo\ and 
\gadget\ codes.  In principle both codes are equally well-suited
to performing dark matter-only calculations (in terms of their
ability to obtain results that both match analytical estimates and
also agree with output from the other code), but practically speaking
the slower speed of the AMR code makes it undesirable as a tool for doing
large, high-resolution N-body calculations at the present day.  
It should be noted that solving Poisson's equation on an adaptive
mesh grid is a relatively new technique, particularly compared
to doing N-body calculations using tree and PM codes, 
and much can be done to speed up the \enzo\ Poisson solver and decrease
its memory consumption.
The \gadget\ N-body solver is already very highly optimized.  If the
speed of the \enzo\ N-body solver can be increased by a factor 
of a few, an improvement which is quite reasonable to expect in the near 
future, the overall speed that the codes require to achieve solutions 
with similar dark matter force resolutions and mass functions will be 
comparable.

In future work it will be important to understand the origin of the
small but finite differences between \enzo/\zeus, \enzo/\ppm,
and SPH at a more fundamental level.  These differences will most
likely be seen (and the reasons for the differences identified) when 
making direct comparisons of the formation and evolution of individual 
dark matter halos and the gas within them.  Additionally, isolated
idealized cases such as the Bertschinger adiabatic infall solution
\citep{Bert85} will provide useful tests to isolate numerical issues.
Examination of individual halos may also point the way to
improved parameterizations of artificial viscosity (and/or
diffusivity) which would then also be beneficial for the SPH
method. Simultaneously, we plan to investigate the differences of the
current generation of codes when additional physical effects such as 
radiative cooling or a UV background are included, and the impact 
of star formation and feedback is modeled.


\section*{Acknowledgements}

This work was supported in part by NSF grants ACI 96-19019, AST 98-02568, 
AST 99-00877, and AST 00-71019. BWO has been funded in part
under the auspices of the U.S.\ Dept.\ of Energy, and supported by its
contract W-7405-ENG-36 to Los Alamos National Laboratory. 
Computational resources were provided by the National 
Center for Supercomputing Applications and the San Diego Supercomputer Center 
under NRAC allocation MCA98N020.
Some SPH simulations were performed at the Center for Parallel 
Astrophysical Computing at Harvard-Smithsonian Center for Astrophysics.
We would like to thank Greg Bryan and Tom Abel for useful discussions, and 
an anonymous referee for suggestions which have greatly improved the 
quality of this manuscript.



\begin{deluxetable}{ccccccc}
\tablecolumns{5}
\tablewidth{0pt}
\tablecaption{\gadget\ Simulations}
\tablehead{\colhead{Run} & \colhead{$\Le$} & \colhead{$N_{\rm part}$} & \colhead{$m_{\rm DM}$} & \colhead{$m_{\rm gas}$} & \colhead{$\epsilon$} & \colhead{notes}}
\startdata
L12N64\_dm & 2048 & $64^3$ & $5.5\times 10^8$ & --- & 5.86 & DM only\\
L12N128\_dm & 3840 & $128^3$ & $6.9\times 10^7$ & --- & 3.13 & DM only\\
L12N256\_dm & 7680 & $256^3$ & $8.6\times 10^6$ & --- & 1.56 & DM only\\
\tableline \\
L3N64\_3.1e & 256 & $2\times 64^3$ & $7.4\times 10^6$ & $1.1\times 10^6$ & 11.7 & Adiabatic \\
L3N64\_3.2e & 512 & $2\times 64^3$ & $7.4\times 10^6$ & $1.1\times 10^6$ & 5.86 & Adiabatic \\
L3N64\_3.3e & 1024 & $2\times 64^3$ & $7.4\times 10^6$ & $1.1\times 10^6$ & 2.93 & Adiabatic \\
L3N64\_3.4e & 2048 & $2\times 64^3$ & $7.4\times 10^6$ & $1.1\times 10^6$ & 1.46 & Adiabatic \\
L3N128 & 3200 & $2\times 128^3$ & $9.3\times 10^5$ & $1.4\times 10^5$ & 0.78 & Adiabatic \\
L3N256 & 6400 & $2\times 256^3$ & $1.2\times 10^5$ & $1.8\times 10^4$ & 0.39 & Adiabatic \\
\enddata
\tablecomments{   
- List of \gadget\ cosmological simulations that are used in 
this study.  $\Le$ is the dynamic range, and $N_{\rm part}$ is 
the particle number 
(in the adiabatic runs there are identical numbers of dark matter 
and gas particles). $m_{\rm DM}$ and $m_{\rm gas}$ are the masses 
of the dark matter and gas particles in units of [$\himsun$]. 
$\epsilon$ is the Plummer-equivalent gravitational softening length 
in units of [$\hikpc$], but the \gadget\ code adopts the spline
kernel. See Section~\ref{gravmethod} for more details.
}
\label{table.sphsim}
\end{deluxetable}

\begin{deluxetable}{ccccc}
\tablecolumns{5}
\tablewidth{0pc}
\tablecaption{\enzo\ Simulations}
\tablehead{\colhead{Run} & \colhead{$\Le$} & \colhead{$N_{\rm DM}$} & \colhead{$\Nroot$} & \colhead{notes}}
\startdata
64g64d\_6l\_dm\_hod & 4096 & $64^3$ & $64^3$ & DM only, high od\\
128g64d\_5l\_dm\_hod & 4096 & $64^3$ & $128^3$ & DM only, high od \\
128g128d\_5l\_dm\_hod & 4096 & $128^3$ & $128^3$ & DM only, high od \\
64g64d\_6l\_dm\_lod & 4096 & $64^3$ & $64^3$ & DM only, low od \\
128g64d\_5l\_dm\_lod & 4096 & $64^3$ & $128^3$ & DM only, low od \\
128g128d\_5l\_dm\_lod & 4096 & $128^3$ & $128^3$ & DM only, low od \\
\tableline \\
64g64d\_6l\_z & 4096 & $64^3$ & $64^3$ & Adiabatic, \zeus \\
64g64d\_6l\_z\_lod & 4096 & $64^3$ & $64^3$ & Adiabatic, \zeus, low OD \\
64g64d\_6l\_q0.5 & 4096 & $64^3$ & $64^3$ & Adiabatic, \zeus, $Q_{\rm AV}=0.5$ \\
128g64d\_5l\_z & 4096 & $64^3$ & $128^3$ & Adiabatic, \zeus \\
128g64d\_5l\_z\_lod & 4096 & $64^3$ & $128^3$ & Adiabatic, \zeus, low OD \\
128g128d\_5l\_z & 4096 & $128^3$ & $128^3$ & Adiabatic, \zeus \\
\tableline  \\
64g64d\_2l\_ppm & 256 & $64^3$ & $64^3$ & Adiabatic, PPM \\
64g64d\_3l\_ppm & 512 & $64^3$ & $64^3$ & Adiabatic, PPM \\
64g64d\_4l\_ppm & 1024 & $64^3$ & $64^3$ & Adiabatic, PPM \\
64g64d\_5l\_ppm & 2048 & $64^3$ & $64^3$ & Adiabatic, PPM \\
64g64d\_6l\_ppm & 4096 & $64^3$ & $64^3$ & Adiabatic, PPM \\
64g64d\_6l\_ppm\_lod & 4096 & $64^3$ & $64^3$ & Adiabatic, PPM, low OD \\
128g64d\_5l\_ppm & 4096 & $64^3$ & $128^3$ & Adiabatic, PPM \\
128g64d\_5l\_ppm\_lod & 4096 & $64^3$ & $128^3$ & Adiabatic, PPM, low OD \\
128g128d\_5l\_ppm & 4096 & $128^3$ & $128^3$ & Adiabatic, PPM \\
128g128d\_5l\_ppm & 4096 & $128^3$ & $128^3$ & Adiabatic, PPM, low OD \\
\enddata
\tablecomments{   
- List of \enzo\ simulations used in this study.
$\Le$ is the dynamic range ($e$ is the size of the finest resolution element,
i.e. the spatial size of the finest level of grids), $N_{\rm
DM}$ is the number of dark matter particles, and $\Nroot$ is the size
of the root grid. `\zeus' and `PPM' in the notes indicate the adopted
hydrodynamic method. `low OD' means that the low overdensity threshold
for refinement were chosen (cells refine with a baryon overdensity
of 4.0/dark matter density of 2.0). `$Q_{\rm AV}$' is the artificial viscosity
parameter for the \zeus\ hydro method when it is not the default value
of 2.0.}
\label{table.amrsim}
\end{deluxetable}


\clearpage

\clearpage   

\end{document}